\documentclass[fleqn,usenatbib]{mnras}

\usepackage{newtxtext,newtxmath}
\usepackage[T1]{fontenc}

\DeclareRobustCommand{\VAN}[3]{#2}
\let\VANthebibliography\thebibliography
\def\thebibliography{\DeclareRobustCommand{\VAN}[3]{##3}\VANthebibliography}

\usepackage{graphicx}
\usepackage{amsmath}

\usepackage{amssymb}
\usepackage{enumerate}

\title[Accretion Flows onto a Neutron Star]{Numerical Studies of Accretion Flows onto a Neutron Star Engulfed in a Massive Star}

\author[D. Sakurai, R. Akaho \& S. Yamada]{
Daiyu Sakurai,$^{1}$\thanks{E-mail: sakurai@heap.phys.waseda.ac.jp}
Ryuichiro Akaho$^{2}$
and Shoichi Yamada$^{3}$
\\
$^{1}$Graduate School of Advanced Science and Engineering, Waseda University, Tokyo 169-8555, Japan\\
$^{2}$Faculty of Science and Engineering, Waseda University, Tokyo 169-8555, Japan\\
$^{3}$Advanced Research Institute for Science and Engineering, Waseda University, 3-4-1 Okubo, Shinjuku, Tokyo 169-8555, Japan
}

\date{Accepted 2026 August 10. Received 2026 August 5; in original form 2026 April 21\\
Published 2026 September 3. doi: \href{https://doi.org/10.1093/mnras/stag1542}{10.1093/mnras/stag1542}}
\pubyear{2026}
\volume{551}

\begin{document}
\label{firstpage}
\pagerange{stag1542}
\maketitle

%-----------------------------------------------------------------
\begin{abstract}
Massive stars commonly form binaries that can evolve into compact systems via common envelope evolution (CEE), a critical but poorly understood phase---especially when the companion is a neutron star. Understanding the force exerted on a neutron star by the surrounding envelope during CEE is a key to the quantitative evaluation of orbital decay, merger time-scale, and compactness of the resultant binary. In this paper, we conduct general-relativistic hydrodynamical simulations under a novel strategy of multi-layer domain-decomposition to treat the vast disparity of $10^4$--$10^7$ between the neutron star radius and the accretion radius. Our 10-model survey spans diverse physical conditions that the neutron star may encounter in the envelope of a massive star. We find that nested bow shocks with alternating orientations commonly form. This configuration is qualitatively different from those in the conventional picture and results in an enhancement of the force by one to two orders of magnitude relative to the Bondi--Hoyle--Lyttleton momentum-accretion estimate. Moreover, some of our models that correspond to the inner region of the massive-star envelope produce positive net projected forces, i.e., thrust rather than drag expected in the standard picture. Although the sign reversal in the net projected force should be regarded as still tentative, these results will serve as a basis for improvements of the force prescription in CEE modelling and have implications for binary evolution theory.
\end{abstract}

%-----------------------------------------------------------------
\section{Introduction} \label{sec:1}

Massive stars are commonly found in binary (or higher-order multiple) systems according to recent spectroscopic and photometric observations \citep{2012Sci...337..444S, 2017IAUS..329..110S, 2024NatAs...8..472L}. Such systems can form compact binaries that host neutron stars (NSs) and black holes (BHs) through orbital evolution, mass transfer and tidal interactions \citep{2014LRR....17....3P, 2024ARA&A..62...21M}. A crucial phase in their evolutionary pathway is the common envelope evolution \citep[CEE;][]{1976IAUS...73...75P}, during which one star engulfs its companion, driving a rapid orbital decay---a process also referred to as spiral-in---as orbital energy is transferred from the companion to the surrounding envelope. If sufficient energy is deposited, the envelope is ejected and a close binary emerges. If envelope ejection fails, the compact object may merge with the stellar core and, depending on the angular momentum that the compact object imparts to the envelope and the subsequent response of the latter, a luminous transient may be produced or a Thorne--\.{Z}ytkow-object-like configuration may be formed \citep[e.g.][]{2025Ap&SS.370...11G,2026enap....3..336O}. If the envelope is ejected, on the other hand, the resultant binary will be compact enough to merge within a Hubble time, becoming a prime source of gravitational waves (GWs) \citep{2000ARA&A..38..113T,2013A&ARv..21...59I, 2020cee..book.....I, 2023LRCA....9....2R}.

The astrophysical importance of CEE has been vindicated by the recent detections of such merging compact binaries by the LIGO/Virgo/KAGRA collaboration \citep{2023PhRvX..13d1039A, 2026ApJ..1005L..51A}. Indeed, reliable theoretical estimations of the formation rates and orbital properties of compact binaries are sensitive to the modelling of CEE. Population-synthesis studies have shown that uncertainties in the CEE treatment---particularly in the orbital decay and envelope ejection---translate into large variations in the merger rate and source population predicted \citep{2012ApJ...759...52D, 2016Natur.534..512B, 2018MNRAS.481.4009V, 2021MNRAS.508.5028B, 2022MNRAS.516.5737B, 2022PhR...955....1M}. A quantitative understanding of the physical processes governing the spiral-in phase is therefore essential for connecting stellar evolution theory to the GW observations.

During spiral-in, the companion experiences strong hydrodynamic drag as it moves through the envelope. This drag transfers energy and angular momentum from the companion to the envelope matter and thus regulates the orbital decay, the envelope ejection, and ultimately the fate of binary \citep{2008ApJ...672L..41R, 2012ApJ...746...74R, 2012ApJ...744...52P}. In the study of systems with a compact object as a companion---such as those with an NS embedded in a red supergiant (RSG)---the drag is commonly estimated by some simplified prescription. The gravitational drag, for example, is inferred from the gravitational attraction of a point mass (a surrogate for the NS) by the matter in its wake \citep{1999ApJ...513..252O, 2007ApJ...665..432K}, while the drag originated from momentum accretion is often modelled with the Bondi--Hoyle--Lyttleton (BHL) formalism \citep{1939PCPS...35..405H, 1944MNRAS.104..273B,1952MNRAS.112..195B}.

Global three-dimensional simulations of CEE have been conducted by several groups so far \citep{2012ApJ...746...74R, 2015MNRAS.450L..39N, 2016ApJ...816L...9O, 2018MNRAS.477.2349I, 2019MNRAS.490.3727C,2020A&A...644A..60S, 2022MNRAS.512.5462L, 2022A&A...667A..72M, 2022A&A...660L...8O}. Related global simulations that include companion-launched jets or common-envelope jets supernovae have also been reported \citep{2019MNRAS.488.5615S,2022MNRAS.514.3212H}. In these calculations, the spatial resolution around the compact object is limited by the extreme dynamic range. Therefore, estimates of the drag and accretion rate are frequently calibrated by local (wind-tunnel) simulations, which have significantly advanced our understanding of the accretion flow and the drag on compact objects in recent years \citep{2015ApJ...803...41M, 2017ApJ...838...56M, 2020MNRAS.497.2057L, 2020ApJ...897..130D, 2020ApJ...894..147C, 2023ApJ...950...31K, 2026arXiv260104188G}. However, because of the extreme spatial dynamic range, such simulations often replace the compact object with an artificial hole that is many orders of magnitude larger than its physical radius. As a result, the flow is not resolved down to the compact object.

To address this issue, we present axisymmetric general-relativistic hydrodynamical (GRHD) simulations of accretion flows onto an NS embedded in an RSG envelope. We work locally in the comoving frame of the NS, in which the envelope gas flows towards the NS as a steady wind. The main difficulty is the extreme contrast between the scale in the vicinity of the NS and that of the RSG envelope, which has prevented seamless physical modelling of the flow from the accretion scale down to the NS surface in previous works.

To overcome this challenge, we adopt a multi-layer domain-decomposition strategy, deploying several consecutively smaller domains from the outside in; these layers overlap at their boundaries and collectively cover the entire computational domain, with each domain sized to resolve the local flow scale. We incorporate consistently general relativity and realistic thermodynamics through the Helmholtz equation of state (EOS).

The goal of this study is to determine the accretion flow from the RSG envelope down to the NS surface, calculate the net projected force exerted on the NS across a wide range of physical conditions encountered during CEE, and thereby provide physically grounded inputs for modelling the spiral-in and eventual formation of compact binaries.

This paper is organized as follows. In Section~\ref{sec:2}, we introduce the physical picture and stellar models that underlie our parameter survey. Section~\ref{sec:3} describes the numerical method. In Section~\ref{sec:4}, we present the main results: the accretion flow structure and the evaluation of accretion rates and net projected forces. Section~\ref{sec:5} discusses the implications for CEE, and we close the paper with our conclusions in Section~\ref{sec:6}.

%-----------------------------------------------------------------
\section{Modelling Accretion Flows in CEE} \label{sec:2}

\begin{figure}
    \centering
    \includegraphics[width=\linewidth]{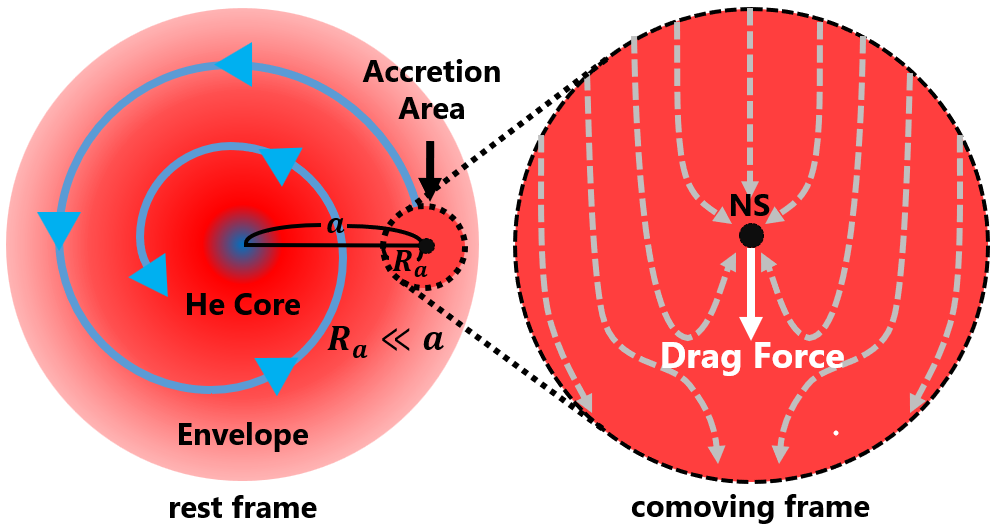}
    \caption{Schematic illustration of an NS engulfed in an RSG envelope during the spiral-in phase. The accretion radius $R_a$ gives the effective cross section of the NS, characterizing the gravitationally-focused flow near the NS. Since $R_a \ll a$ (orbital separation), the flow can be modelled locally by using the physical conditions of the RSG envelope as the boundary condition.}
    \label{fig:ns_rsg_schematic}
\end{figure}

In this section, we outline the physical picture to model accretion flows onto an NS embedded in the common envelope of an RSG (Fig.~\ref{fig:ns_rsg_schematic}). We adopt a local description in the comoving frame of the NS, in which the RSG envelope gas flows past the NS as a steady wind. Such a picture is motivated by the short flow-crossing time across the accretion radius compared with the local orbital period, as estimated below. We first summarize the relevant parameters that characterize the steady flows (Section~\ref{sec:2.1}), and then describe the stellar models that give the boundary conditions in our simulations (Section~\ref{sec:2.2}). Finally, we discuss the limitations of standard approaches to motivate our multi-layer numerical strategy (Section~\ref{sec:2.3}).

\subsection{Characteristic Parameters of Accretion Flows} \label{sec:2.1}

As a reference, we adopt the BHL description of accretion flows \citep{2004NewAR..48..843E}, in which an NS of mass $M_\mathrm{NS}$ moves through a uniform medium of density $\rho_\infty$ and sound speed $c_s$ at a velocity $v_\infty$. Gravitational focusing defines the accretion radius as
\begin{equation}
    R_a \equiv \frac{2 G M_\mathrm{NS}}{v_\infty^2 + c_s^2},
    \label{eq:Ra_def}
\end{equation}
which gives the spatial scale over which the flow is significantly perturbed. Here and throughout this paper, $G$ denotes the gravitational constant. The mass accretion rate, drag force, and energy deposition rate are estimated in this model as
\begin{align}
    \dot{M}_\mathrm{BHL} &= \pi R_a^2 \rho_\infty v_\infty, \label{eq:mdot_bhl} \\
    F_\mathrm{BHL} &= \dot{M}_\mathrm{BHL} v_\infty, \label{eq:f_bhl} \\
    \dot{E}_\mathrm{BHL} &= F_\mathrm{BHL} v_\infty, \label{eq:edot_bhl}
\end{align}
respectively. The physical force vector points opposite to the motion. We note that $F_\mathrm{BHL}$ accounts only for the momentum carried by the accreted matter and does not include the gravitational drag (dynamical-friction) exerted by matter in the wake \citep[e.g.,][]{1999ApJ...513..252O}. Nonetheless, we use it as a reference in the comparison with the forces that we obtain in the simulations, including the gravitational drag. In so doing, we refer to $F_\mathrm{BHL}$ always as the BHL momentum-accretion estimate to make clear that it is based on the accreted momentum alone and the gravitational force is excluded. This is because the analytical estimate of the gravitational drag depends on assumptions about the Coulomb logarithm and the wake that are not well constrained for the conditions of interest. In this paper we evaluate all components contributing to the drag directly from our simulation results.

For the NS spiralling in the RSG envelope, the incoming velocity of matter is approximately given by the local Kepler velocity,
\begin{equation}
    v_\infty \approx v_\mathrm{kep}(a)
    \equiv
    \sqrt{\frac{G m_\mathrm{RSG}(a)}{a}},
    \label{eq:vinf_kep}
\end{equation}
where $m_\mathrm{RSG}(a)$ is the enclosed mass of the RSG up to the radius $a$. Hereafter, we refer to $a$ (the distance between the centers of NS and RSG) as the orbital separation. The accretion radius is typically much smaller than the orbital separation, $R_a \ll a$. This may be confirmed by substituting Eq.~(\ref{eq:vinf_kep}) into Eq.~(\ref{eq:Ra_def}) and using $\mathcal{M}\equiv v_\infty/c_s$ as follows,
\begin{equation}
    \frac{R_a}{a}
    \approx
    \frac{2M_\mathrm{NS}}{m_\mathrm{RSG}(a)}
    \frac{\mathcal{M}^2}{1+\mathcal{M}^2}.
    \label{eq:Ra_over_a}
\end{equation}
For an NS moving through a massive-star envelope, $\mathcal{M}\sim1$--2 and $M_\mathrm{NS}/m_\mathrm{RSG}(a) \sim 10^{-2}$--$10^{-1}$. Then Eq.~(\ref{eq:Ra_over_a}) gives $R_a/a\sim10^{-2}$--$10^{-1}$. Hence the accretion flow may be treated locally, that is, we isolate a region around the NS on the scale of the accretion radius from the RSG envelope. On the other hand, the flow-crossing time,
$t_\mathrm{cross} \equiv R_a / v_\infty$, may be compared with the local orbital period, $t_\mathrm{orb}\equiv 2\pi a/v_\mathrm{kep}$, as
\begin{equation}
    \frac{t_\mathrm{cross}}{t_\mathrm{orb}}
    \approx
    \frac{R_a}{2\pi a}
    \sim 10^{-3}\text{--}10^{-2}.
    \label{eq:tcross_orb}
\end{equation}
This justifies the quasi-steady approximation adopted here \citep{2015ApJ...803...41M}. In this local model, we neglect the Coriolis force and the residual of the cancellation between the centrifugal and gravitational forces, and we also ignore the density gradient in the incoming flow. The validity of the latter assumption can be quantified by the dimensionless parameter $\varepsilon_\rho \equiv R_a/H_\rho$, where $H_\rho$ is the density scaleheight; when $\varepsilon_\rho\ll1$, the density stratification across the accretion radius can be ignored. With these simplifications, the local flow is approximated as axisymmetric with respect to the direction of NS motion, i.e., the z-direction in this study.

\subsection{Stellar Models} \label{sec:2.2}

In our local simulations of the accretion flow, the information about the RSG envelope is encoded in the outer boundary conditions. We obtain the RSG envelope structures using the stellar evolution code MESA \citep{2019ApJS..243...10P}. By sampling the density and temperature at various radii (representing different NS locations), we adopt these values as $\rho_\infty$ and $T_\infty$ for the incident matter at the outer boundary. The velocity $v_\infty$ is given by the local Kepler velocity defined in Eq.~(\ref{eq:vinf_kep}).

\begin{figure}
    \centering
    \includegraphics[width=\linewidth]{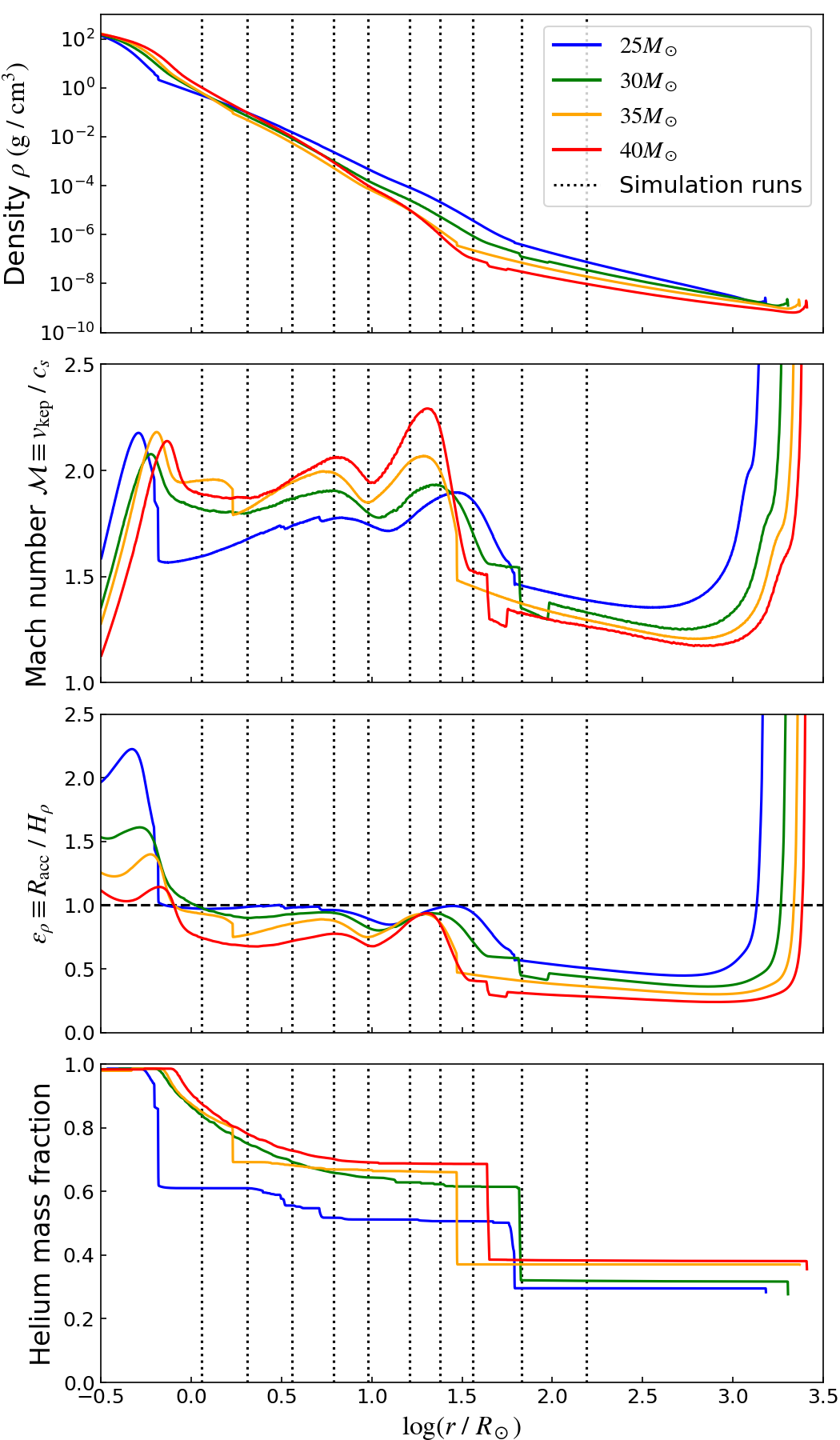}
    \caption{Radial structure of RSG envelopes at the onset of core helium burning, computed with MESA, for zero-age main-sequence masses of 25, 30, 35, and $40\,M_\odot$. Shown from top to bottom are the density, Mach number, dimensionless parameter, and helium mass fraction; vertical dotted lines mark the radii at which the data are sampled to build the models listed in Table~\ref{tab:twocol1}.
    }
    \label{fig:mesa_structure}
\end{figure}

Figure~\ref{fig:mesa_structure} presents the radial profiles of density $\rho$, Mach number $\mathcal{M}$, dimensionless parameter $\varepsilon_\rho$, and helium mass fraction for the RSG models with zero-age main-sequence masses of 25, 30, 35, and $40\,M_\odot$, evaluated at the onset of core helium burning. The density changes by $\sim$10 orders of magnitude, and the NS moves at supersonic speeds ($\mathcal{M}>1$) across much of the envelope. Among these models, we use the last one with $M_{\rm ZAMS}=40\,M_\odot$, whose total mass is $M_{\rm RSG}=32.8\,M_\odot$ at the RSG phase, because it has the widest radial interval with $\varepsilon_\rho<1$. At the sampled radii (vertical dotted lines in Fig.~\ref{fig:mesa_structure}), $\varepsilon_\rho$ is smaller than unity, indeed supporting our neglect of the density gradient in the incoming flow (see Section~\ref{sec:2.1}). For less massive models and/or near the stellar surface or core, $\varepsilon_\rho$ exceeds unity (Fig.~\ref{fig:mesa_structure}), and the density gradient cannot be ignored; it may influence the gas flow and, consequently, the force exerted on the NS \citep{2017ApJ...838...56M,2020ApJ...897..130D,2026arXiv260104188G}. Then 3D simulations are required. We defer them to future work, though.

\begin{table*}
\centering
\caption{Parameters that characterize the numerical models. From the left, the columns give the model name, the logarithmic orbital radius $\log(a/R_\odot)$, the density $\rho_\infty$, temperature $T_\infty$, pressure $P_\infty$, velocity $v_\infty$, Mach number $\mathcal{M}_\infty$, adiabatic index $\gamma_\infty$, and mass fractions of $^1$H, $^4$He, and $^{12}$C in the incoming flow and the EOS. The first three models are controlled test calculations with idealized conditions chosen to isolate the flow physics, whereas the next ten models use the physical conditions sampled at different radii (vertical dotted lines in Fig.~\ref{fig:mesa_structure}) in the envelope of the $40\,M_\odot$ RSG model provided by MESA.}
\label{tab:twocol1}
\begin{tabular}{cccccccccccc}
\hline
model & $\log (a / R_\odot)$ & $\rho_\mathrm{\infty} (\mathrm{g/cm^3})$ & $T_\mathrm{\infty} (\mathrm{K})$ & $P_\mathrm{\infty} (\mathrm{erg/cm^3})$ & $v_\mathrm{\infty} (\mathrm{cm/s})$ & $\mathcal{M}_\mathrm{\infty}$ & $\gamma_\mathrm{\infty}$ & $^1\mathrm{H}$ & $^4\mathrm{He}$ & $^{12}\mathrm{C}$ & EOS \\
\hline
\texttt{test-Heos}        & - & $10^{3}   $ & $4.6  \times 10^{8}$ & $1.8 \times 10^{20}$ & $1   \times 10^{9}$ & 2   & 1.33 & 0.75  & 0.23  & 0.02  & Helmholtz \\
\texttt{test-$\gamma$5/3} & - & $10^{3}   $ & -                    & $1.5 \times 10^{20}$ & $1   \times 10^{9}$ & 2   & 1.67 & -     & -     & -     & gamma-law \\
\texttt{test-$\gamma$4/3} & - & $10^{3}   $ & -                    & $1.9 \times 10^{20}$ & $1   \times 10^{9}$ & 2   & 1.33 & -     & -     & -     & gamma-law \\
\texttt{$\rho$1e0}  &  $0.06$ & $1        $ & $3.5  \times 10^{7}$ & $6.3 \times 10^{15}$ & $1.8 \times 10^{8}$ & 2.0 & 1.41 & 0.111 & 0.875 & 0.014 & Helmholtz \\
\texttt{$\rho$1e-1} &  $0.31$ & $10^{-1}  $ & $1.8  \times 10^{7}$ & $3.8 \times 10^{14}$ & $1.4 \times 10^{8}$ & 1.9 & 1.40 & 0.205 & 0.781 & 0.014 & Helmholtz \\
\texttt{$\rho$1e-2} &  $0.56$ & $10^{-2}  $ & $8.5  \times 10^{6}$ & $2.0 \times 10^{13}$ & $1.1 \times 10^{8}$ & 2.0 & 1.40 & 0.257 & 0.729 & 0.014 & Helmholtz \\
\texttt{$\rho$1e-3} &  $0.79$ & $10^{-3}  $ & $4.2  \times 10^{6}$ & $1.1 \times 10^{12}$ & $8.5 \times 10^{7}$ & 2.1 & 1.39 & 0.282 & 0.704 & 0.014 & Helmholtz \\
\texttt{$\rho$1e-4} &  $0.98$ & $10^{-4}  $ & $2.2  \times 10^{6}$ & $8.0 \times 10^{10}$ & $6.8 \times 10^{7}$ & 2.0 & 1.38 & 0.295 & 0.691 & 0.014 & Helmholtz \\
\texttt{$\rho$1e-5} &  $1.21$ & $10^{-5}  $ & $1.0  \times 10^{6}$ & $3.6 \times 10^{9 }$ & $5.2 \times 10^{7}$ & 2.3 & 1.38 & 0.297 & 0.689 & 0.014 & Helmholtz \\
\texttt{$\rho$1e-6} &  $1.38$ & $10^{-6}  $ & $5.4  \times 10^{5}$ & $2.6 \times 10^{8 }$ & $4.3 \times 10^{7}$ & 2.3 & 1.37 & 0.299 & 0.687 & 0.014 & Helmholtz \\
\texttt{$\rho$1e-7} &  $1.56$ & $10^{-7}  $ & $3.4  \times 10^{5}$ & $3.4 \times 10^{7 }$ & $3.5 \times 10^{7}$ & 1.6 & 1.35 & 0.300 & 0.686 & 0.014 & Helmholtz \\
\texttt{$\rho$3e-8} &  $1.83$ & $10^{-7.5}$ & $2.3  \times 10^{5}$ & $7.9 \times 10^{6 }$ & $2.6 \times 10^{7}$ & 1.4 & 1.35 & 0.601 & 0.385 & 0.014 & Helmholtz \\
\texttt{$\rho$1e-8} &  $2.19$ & $10^{-8}  $ & $1.4  \times 10^{5}$ & $1.2 \times 10^{6 }$ & $2.4 \times 10^{7}$ & 1.3 & 1.36 & 0.603 & 0.383 & 0.014 & Helmholtz \\
\hline
\end{tabular}
\end{table*}

Table~\ref{tab:twocol1} presents all the models studied in this paper. The columns specify the parameters that characterize the models. The first three rows are the controlled test models, \texttt{test-Heos}, \texttt{test-$\gamma$5/3}, and \texttt{test-$\gamma$4/3}. They do not represent particular locations in the RSG envelope. Instead, they are reference cases meant to examine the basic flow features and their EOS-dependence under controlled conditions; their computational setups and results are described in Sections~\ref{sec:3.2} and \ref{sec:4.1}, respectively. As their names indicate, \texttt{test-Heos} uses the Helmholtz EOS with hydrogen-rich matter, whereas \texttt{test-$\gamma$5/3} and \texttt{test-$\gamma$4/3} use the gamma-law EOSs with the corresponding values of $\gamma$. The remaining 10 rows list the MESA-based models, which use the envelope conditions sampled from the $40\,M_\odot$ progenitor model. They are meant to survey a broad range of physical conditions encountered by NS during its spiral-in. These realistic models are named by the density in the incoming flow adopted: for example, \texttt{$\rho$1e-3} corresponds to $\rho_\infty=10^{-3}\,\mathrm{g\,cm^{-3}}$. Since a donor with $M_{\rm ZAMS}=40\,M_\odot$ is more likely to leave a BH than an NS, the parameter regime explored here may be more relevant to the NS--BH or BH--BH progenitor systems than to the NS--NS systems.

We also use the stellar models to evaluate the gravitational binding energy of the envelope and the orbital energy budget relevant for envelope ejection in CEE. These two quantities provide the standard starting point for estimating the post-CE orbital separation. In the conventional $\alpha$-formalism \citep{1984ApJ...277..355W, 1990ApJ...358..189D, 2000A&A...360.1043D}, the gravitational binding energy of the envelope outside a given radius $r$ (measured from the RSG centre) is written as
\begin{equation}
    E_{\rm bind}(r) \equiv \int_{m(r)}^{M_{\rm RSG}}\left(-\frac{G m^\prime}{r^\prime} + \alpha_{\rm th}\,u^\prime \right)\,\mathrm{d}m^\prime,
    \label{eq:Ebind_def}
\end{equation}
where $m(r)$ is the enclosed mass up to $r$; $M_{\rm RSG}$ is the RSG mass; $r^\prime\equiv r(m^\prime)$ is the radius at mass coordinate $m^\prime$; $u^\prime$ is the specific internal energy; $\alpha_{\rm th}$ is a parameter that measures the fraction of thermal energy available to unbind the envelope \citep{2022A&A...667A..72M}. The change of the orbital energy associated with the inspiral of NS from the initial separation $a_i$ to the current one $r$ is given by
\begin{equation}
    \begin{split}
    \Delta E_{\rm orb}(r) &\equiv -\frac{G m(r) M_{\rm NS}}{2r} + \frac{G M_{\rm RSG} M_{\rm NS}}{2a_i}\\
    &\approx -\frac{G m(r) M_{\rm NS}}{2r},
    \end{split}
    \label{eq:DEorb_def}
\end{equation}
where the last approximation holds when $a_i\gg r$.

\begin{figure}
    \centering
    \includegraphics[width=\linewidth]{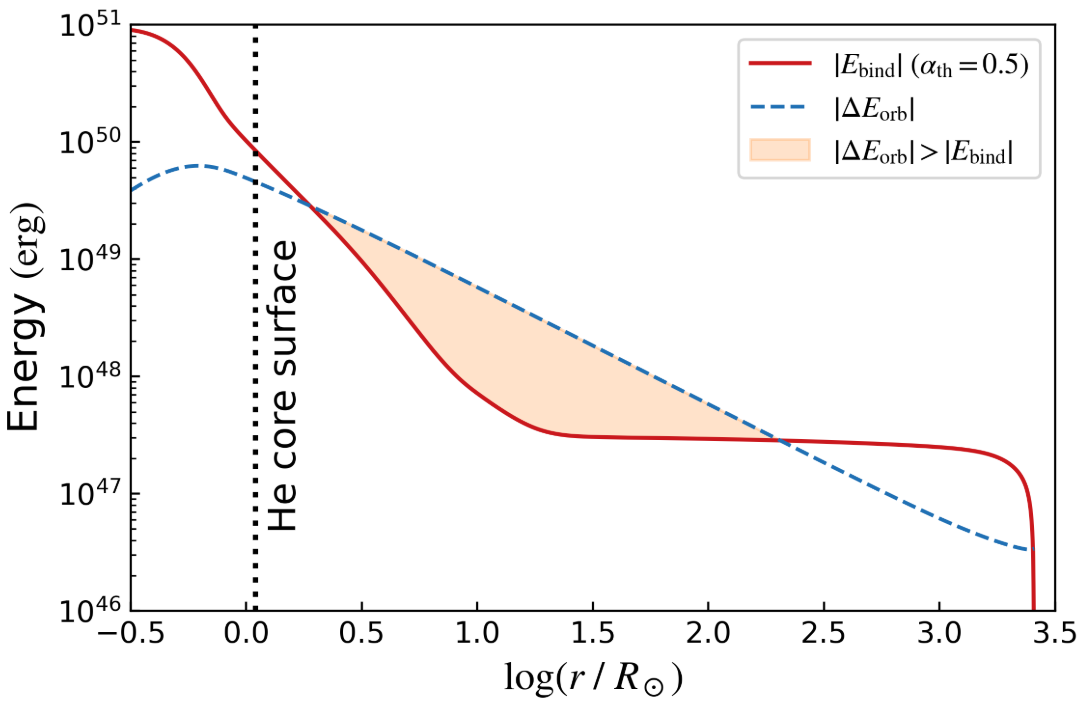}
    \caption{
        Radial profiles of \(|E_{\rm bind}|\) and \(|\Delta E_{\rm orb}|\) for the progenitor model with \(M_{\rm ZAMS}=40\,M_\odot\). The binding energy is evaluated for \(\alpha_{\rm th}=0.5\). The orange-shaded region marks the radial range, where \(|\Delta E_{\rm orb}|>|E_{\rm bind}|\) for \(\alpha_{\rm CE}=1\). The vertical dotted line indicates the He-core surface.
    }
    \label{fig:energy_formalism}
\end{figure}

In the $\alpha$-formalism, the envelope exterior to the final orbital separation, $a_f$, is assumed to be ejected, and $a_f$ is determined as a function of $\alpha_{\rm th}$ and $\alpha_{\rm CE}$ by the following equation:
\begin{equation}
    E_{\rm bind}(a_f) = \alpha_{\rm CE}\,\Delta E_{\rm orb}(a_f),
    \label{eq:alpha_formalism}
\end{equation}
where $\alpha_{\rm CE}$ is the ejection efficiency. Figure~\ref{fig:energy_formalism} shows an exemplary comparison between $|E_{\rm bind}|$, evaluated for $\alpha_{\rm th}=0.5$ and $|\Delta E_{\rm orb}|$ in our progenitor model with $M_{\rm ZAMS}=40\,M_\odot$. The orange-shaded region marks the radial range, $\log(r/R_\odot)\simeq0.28$--2.31, where $|\Delta E_{\rm orb}|>|E_{\rm bind}|$; in the $\alpha$-formalism estimate with $\alpha_{\rm CE}=1$, the envelope exterior to these radii can be unbound.

This energy criterion should be regarded only as an order-of-magnitude guide. The actual outcome can depend on the envelope response, energy losses through radiation and outflows, the location of energy deposition, and the values adopted for $\alpha_{\rm CE}$ and $\alpha_{\rm th}$. Therefore, quantitative simulations are required to evaluate the gas force exerted on the NS, the primary driver of the spiral-in dynamics (see Section~\ref{sec:5.1}).

\subsection{Limitations of Standard Approaches} \label{sec:2.3}

Most previous simulations of accretion flows onto compact objects during CEE employ simplified prescriptions to avoid directly handling the vicinity of a compact object. In practice, the vast scale separation between the NS radius and the accretion radius presents a formidable numerical challenge. As a result, many authors adopt a strategy to replace the compact object with an absorbing boundary that is many orders of magnitude larger than its actual physical scale \citep{2015ApJ...803...41M, 2017ApJ...838...56M, 2020MNRAS.497.2057L, 2020ApJ...897..130D, 2026arXiv260104188G}. Others choose to artificially enhance the accretion velocity so that the accretion radius and therefore the computational domain can be reduced \citep{2020ApJ...894..147C, 2023ApJ...950...31K}. 

In addition, the common use of Newtonian gravity and idealized equations of state in the literature limits the ability to accurately model the flows in the vicinity of a compact object. Consequently, current estimates of the hydrodynamic drag on the NS remain highly uncertain. In order to address this issue, we need to consistently account for general relativistic gravity, realistic thermodynamics, and a wide range of spatial scales. In the following section, we describe the numerical methods we develop for this purpose.

%-----------------------------------------------------------------
\section{Numerical Methods} \label{sec:3}

We perform general relativistic hydrodynamical (GRHD) simulations of axisymmetric accretion flows onto an NS embedded in an RSG envelope during CEE. In order to cover the vast dynamic range involved, we employ a multi-layer domain-decomposition technique: the computational domain is divided into successively narrower concentric layers from the outside in. Adjacent layers overlap with each other by roughly half of their radial widths to facilitate layer transitions. The width of each layer is carefully set to ensure that the local density scaleheight is well resolved. We then solve the hydrodynamical equations consecutively and iteratively across these layers. The iteration continues until the flows in all overlapping regions agree, yielding a globally steady flow across the entire domain. Detailed procedures are provided in Section~\ref{sec:3.2}. We evaluate the mass accretion rate, net projected force, and energy deposition rate based on the resultant flow, as defined in Section~\ref{sec:3.3}.

\subsection{Basic equations} \label{sec:3.1}

Before proceeding to the detailed explanation of the multi-layer domain-decomposition method, we summarize the hydrodynamical equations, equation of state (EOS) and the numerical scheme adopted. The computational domain is centred on and comoving with the NS. The domain size is taken to be 10 times the accretion radius (Eq.~(\ref{eq:Ra_def})) at the given position in the RSG envelope (see Table~\ref{tab:twocol1}). Throughout this paper, Latin indices $i$, $j$, and $k$ run over the spatial components 1--3, and $c$ denotes the speed of light.

We solve the GRHD equations in axisymmetry with the spacetime metric fixed to the Schwarzschild metric for an NS of mass $M_\mathrm{NS}=1.41\,M_\odot$:
\begin{equation}
\begin{split}
ds^2 = &-\left(1-\frac{2GM_\mathrm{NS}}{rc^2}\right)c^2dt^2
+ \left(1-\frac{2GM_\mathrm{NS}}{rc^2}\right)^{-1}dr^2 \\
&+ r^2 d\theta^2 + r^2 \sin^2\theta\, d\phi^2.
\end{split}
\end{equation}
In the $3+1$ form, the shift vector vanishes ($\beta^i=0$) and the lapse function and the spatial metric are given, respectively, as
\begin{align}
\alpha&=\left(1-\frac{2GM_\mathrm{NS}}{rc^2}\right)^{1/2}, \\
\gamma_{ij}&=\mathrm{diag}\!\left[\left(1-\frac{2GM_\mathrm{NS}}{rc^2}\right)^{-1},\, r^2,\, r^2\sin^2\theta\right].
\end{align}
These quantities and $\sqrt{\gamma}\equiv\sqrt{\det(\gamma_{ij})}$ enter the GRHD equations as shown below. Assuming that the centrifugal force is completely canceled by the gravitational force of the RSG, we neglect the self-gravity of the RSG. The Coriolis force is also ignored and therefore the assumption of axisymmetry follows. It is further noted that the envelope-gas mass within the computational domain, ${\sim}\,\rho_\infty(10R_a)^3 \sim 10^{-3}$--$10^{-1}\,M_{\rm NS}$, is negligible compared with the NS mass for our envelope conditions.

We evolve the conservation equations for mass, momentum, and energy in conservative form \citep{2016nure.book.....S}:
\begin{align}
&\partial_t \rho_* + \partial_j(\rho_* v^j) = 0,
\label{eq_massconserv}\\[4pt]
&\partial_t S_i + \partial_j \left(S_i v^j + \alpha\sqrt{\gamma}P c^2\delta_i{}^j\right)
\nonumber\\
&\qquad = -S_0 c^2\partial_i \alpha + S_j c\partial_i \beta^j
- \frac{1}{2}\alpha c^2 \sqrt{\gamma}\, S_{jk}\partial_i \gamma^{jk},
\label{eq_Euler}\\[4pt]
&\partial_t (S_0 - \rho_*c^2) + \partial_k \left((S_0 - \rho_*c^2)v^k + \sqrt{\gamma}P (v^k + c\beta^k)\right)
\nonumber\\
&\qquad = \alpha\sqrt{\gamma}S^{ij}K_{ij} - S_i D^i \alpha.
\label{eq_energyconserv}
\end{align}
where $\rho_*$, $v^j \equiv u^j/u^t$, $S_i$, $S_0$ and $S_{ij}$ are the conserved rest-mass density, three velocity, the momentum and energy densities and the stress tensor, respectively. Their definitions are given as follows:
\begin{align}
\rho_* &\equiv \alpha \sqrt{\gamma}\rho u^t = \sqrt{\gamma}\rho w, \\
S_j &\equiv \rho_* h u_j c,\\
S_0 &\equiv \sqrt{\gamma}(\rho h w^2 - P), \\
w &\equiv \alpha u^t, \\
S_{ij} &\equiv \rho h u_i u_j + P\gamma_{ij},
\end{align}
with $\rho$ the rest-mass density, $P$ the pressure, $u^\mu$ the four-velocity, and $h$ the specific enthalpy.

We adopt the Helmholtz EOS \citep{2000ApJS..126..501T}, which includes radiation, ions, and electron--positron pairs as constituents and is suitable for the wide range of density and temperature (see Table~\ref{tab:twocol1}). Since the RSG envelope considered here is optically thick, we do not solve radiative transfer explicitly but radiation pressure and energy are included through the Helmholtz EOS.

We employ a finite-volume scheme with the MUSCL reconstruction and the HLL Riemann solver. The third-order TVD Runge--Kutta method is adopted for time-integration with a Courant--Friedrichs--Lewy (CFL) number of 0.4. For more details, see \citet{2023ApJ...944...60A}.

\subsection{Multi-layer domain-decomposition} \label{sec:3.2}

The numerical difficulty arises from the very broad dynamic ranges in the length- and time-scales: the accretion radius changes by a factor of $10^4$--$10^7$ and the sound-crossing time varies from $\sim 10^{-4}$\,sec near the NS to $\sim 10^7$\,sec near the RSG surface. Therefore, the simulation on a single mesh that covers the entire computational domain would be prohibitively expensive.

\begin{figure}
    \centering
    \includegraphics[width=1.0\linewidth]{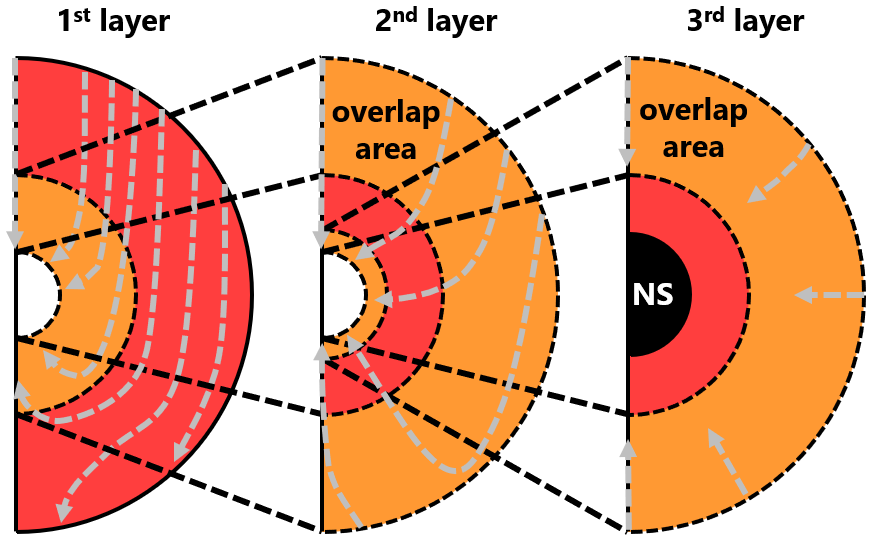}
    \caption{Schematic illustration of the multi-layer domain-decomposition strategy ($n_\mathrm{layer}=3$) used to resolve the multiscale accretion flow. The computational domain is divided into concentric spherical shells. The adjacent layers are partially overlapped (red regions). The hydrodynamical equations are solved in each layer individually and consecutively with the boundary conditions given by the solutions in the adjacent layers. The procedure is continued until the global flow (grey dashed lines) is obtained consistently in the entire computational domain.}
    \label{fig:multi-layer}
\end{figure}

We adopt a multi-layer domain-decomposition strategy. We divide the computational domain into $n_\mathrm{layer}=2$--3 concentric spherical shells, each spanning $\sim$2--3 decades in radius. As shown schematically in Fig.~\ref{fig:multi-layer}, the neighbouring layers are overlapped with each other at their boundaries. We want to solve the hydrodynamical equations in each layer individually so that the resultant flows should agree with each other in these overlapping regions. The key idea to accomplish this is a bidirectional iteration between adjacent domains. The concrete procedure is as follows.

We start with the outermost layer (the 1st layer in Fig.~\ref{fig:multi-layer}). We solve the hydrodynamical equations until a steady state is obtained. The initial condition is rather arbitrary but we take a uniform flow in the negative $z$ direction with the density and temperature equal to those of the RSG envelope at the position of the NS. The flow speed is set to the local Kepler velocity at that position (see Table~\ref{tab:twocol1}). We deploy the spherical coordinates with the NS at the centre. As mentioned already, we assume that the system is axisymmetric with respect to the $z$ axis and therefore the matter profile is independent of $\phi$. As the boundary conditions, we impose on the upper hemisphere $(0 \leq \theta \leq \pi/2)$ the Dirichlet-type, i.e., fixed inflow condition with the same density, temperature and velocity as those in the initial condition while we employ the Neumann-type, i.e., zero-gradient outflow condition on the lower hemisphere.

Once the steady solution is obtained in the outermost layer, we move on to the next inner layer (see Fig.~\ref{fig:multi-layer}). This time the outer boundary condition is given by the result of the previous step. The initial condition is the same as that for the first step but with the velocity vanishing. Since we are interested in the steady solution, the initial condition is not very important unless it is very different from the final result. As for the inner boundary condition we impose again the Neumann-type, i.e., zero-gradient condition if there is another layer further inside. If there are only two layers, the inner boundary of the second layer corresponds to the NS surface. We impose the zero-gradient condition for the thermodynamic variables. For the velocity, on the other hand, we set $\boldsymbol{v} = 0$, modelling the accreting matter brought to rest at the NS surface. It is demonstrated in Appendix~\ref{sec:d} that the estimation of the net projected force is insensitive to reasonable variations of the inner boundary condition.

In general the steady solution so obtained in the second layer does not agree in the overlapping region with the steady solution derived in the first step for the first layer. We then go back to the first layer and redo the simulation with the inner boundary condition updated by the solution in the second layer. The updated solution in the first layer does not agree in the overlapping layer with the solution in the second layer. Then we update the latter solution with the outer boundary condition updated by the updated solution in the first layer. This iteration is repeated until the satisfactory agreement in the overlapping region is obtained between the steady solutions for the two layers.

If there is a third layer, we proceed to it after consistency between the first and second layers has been achieved. The inner boundary of the third layer is the NS surface in this case. We impose the Dirichlet-type condition there just as for the inner boundary of the second layer when there are only two layers. The outer boundary condition is given by the steady solution obtained after the previous steps. The third-layer calculation is then continued until the flow settles into a quasi-steady state.

In the inner layers, the accretion flow often does not reach a perfectly steady state; instead, shock oscillations persist. To prevent these time variations from affecting the global matching, we carefully place the overlapping regions sufficiently far from the unsteady regions. This is always possible, allowing the consistency condition to be imposed unambiguously. From our experience, this consistency is rather easily obtained without further iterations for inner regions if the consistency between the first two layers has been achieved. If necessary, however, we return to the first layer and repeat the above steps.

In practice, we require for the consistency that physical quantities (specifically density and temperature) should agree between the two solutions within $\sim 0.1\%$ in the overlapping region over $\sim 10$ dynamical times. This criterion is sufficient in practice to recover the same global solution, including the force and flux diagnostics, because the overlapping regions are placed in smooth, quasi-steady parts of the flow away from the oscillatory inner zone, as confirmed by the convergence tests in Appendix~\ref{sec:a}. After convergence is achieved for a given pair, we proceed inward to the next domain. Appendix~\ref{sec:a} presents the technical details and convergence tests.

Our method is similar to the Schwarz alternating method originally proposed for elliptic partial differential equations \citep{Schwarz1870,Lions1988,Quarteroni1999,Toselli2005}. Exploiting the steady nature of the accretion flow of our concern, we apply the analogous idea to the hyperbolic differential equations. We note that a similar multi-zone domain-decomposition strategy was employed recently for the simulation of accretion flows in active galactic nuclei (AGN) \citep{2024ApJ...977..200C}.

Finally, we describe the numerical grid employed in our simulations. Within each layer, the radial grid is logarithmically spaced. The innermost layer starts at the NS surface ($R_\mathrm{NS} = 1.15 \times 10^6\,\mathrm{cm}$), where the minimum radial grid size is set to $\Delta r \approx 10^4\,\mathrm{cm}$. Moving outward, the grid spacing gradually increases by approximately $5\%$ per cell, ensuring a smooth transition in resolution up to the outer boundary at $r = 10 R_a$. The polar grid covers the full range $0 \le \theta \le \pi$ with Gauss--Legendre quadrature nodes in $\mu = \cos\theta$, using $N_\theta = 48$ grid points. The radial grid points are chosen so that they coincide exactly in the overlapping region for every pair of adjacent layers, facilitating stable data exchanges.

Table~\ref{tab:layers} lists the layer structure and grid configuration adopted for each model. The realistic models generally have larger accretion radii than the test models and hence require two or three layers ($n_{\rm layer}=2,3$) to resolve the flow from the NS surface to the outer boundary with reasonable resolutions. In contrast, the three test models are designed so that they could have sufficiently small accretion radii to be evolved in a single layer ($n_{\rm layer}=1$), making them useful for controlled studies of the shock morphology. Indeed, Appendix~\ref{sec:a} validates the multi-layer procedure in the \texttt{test-Heos} configuration by comparing the one-layer computation with another simulation with two layers, which are not necessary but intentionally deployed.

\begin{table*}
\centering
\caption{Configurations of the computational domain: layer structure and grid resolutions. The number of radial cells in the $i$th layer and that across the full computational domain are denoted by $N_r^{i^\mathrm{th}}$ and $N_r^{\mathrm{all}}$, respectively. Owing to the overlap between neighbouring layers, the latter is not equal to the sum of $N_r^{i^\mathrm{th}}$ over all layers.}
\label{tab:layers}
\begin{tabular}{ccccccccccc}
\hline
model & $n_\mathrm{layer}$ &
$R_\mathrm{in}^{3^\mathrm{rd}} \text{--} R_\mathrm{out}^{3^\mathrm{rd}}\,(\mathrm{cm})$ & $N_r^{3^\mathrm{rd}}$ &
$R_\mathrm{in}^{2^\mathrm{nd}} \text{--} R_\mathrm{out}^{2^\mathrm{nd}}\,(\mathrm{cm})$ & $N_r^{2^\mathrm{nd}}$ &
$R_\mathrm{in}^{1^\mathrm{st}} \text{--} R_\mathrm{out}^{1^\mathrm{st}}\,(\mathrm{cm})$ & $N_r^{1^\mathrm{st}}$ & $N_r^{\mathrm{all}}$ &
$N_{\theta}$ \\
\hline
\texttt{test-Heos}        & 1 & $1.15 \times 10^{6} \text{--} 3.0 \times 10^{9}$ & $168$ & $-$ & $-$ & $-$ & $-$ & $168$ & $48$ \\
\texttt{test-$\gamma$5/3} & 1 & $1.15 \times 10^{6} \text{--} 3.0 \times 10^{9}$ & $168$ & $-$ & $-$ & $-$ & $-$ & $168$ & $48$ \\
\texttt{test-$\gamma$4/3} & 1 & $1.15 \times 10^{6} \text{--} 3.0 \times 10^{9}$ & $168$ & $-$ & $-$ & $-$ & $-$ & $168$ & $48$ \\
\texttt{$\rho$1e0}        & 2 &
 $1.15 \times 10^{6} \text{--} 1.5 \times 10^{9}$ & $168$ & 
 $7.5\times10^{8} \text{--} 8.8\times10^{10}$ & $98$ & $-$ & $-$ & $252$ & $48$ \\
\texttt{$\rho$1e-1}       & 2 &
 $1.15 \times 10^{6} \text{--} 2.0 \times 10^{9}$ & $168$ & 
 $1.0\times10^{9} \text{--} 1.4\times10^{11}$ & $98$ & $-$ & $-$ & $252$ & $48$ \\
\texttt{$\rho$1e-2}       & 2 &
 $1.15 \times 10^{6} \text{--} 2.9 \times 10^{9}$ & $168$ & 
 $1.4\times10^{9} \text{--} 2.5\times10^{11}$ & $98$ & $-$ & $-$ & $252$ & $48$ \\
\texttt{$\rho$1e-3}       & 2 &
 $1.15 \times 10^{6} \text{--} 3.3 \times 10^{9}$ & $182$ & 
 $1.7\times10^{9} \text{--} 4.2\times10^{11}$ & $112$ & $-$ & $-$ & $280$ & $48$ \\
\texttt{$\rho$1e-4}       & 2 &
 $1.15 \times 10^{6} \text{--} 2.1 \times 10^{9}$ & $168$ & 
 $1.0\times10^{9} \text{--} 6.5\times10^{11}$ & $126$ & $-$ & $-$ & $280$ & $48$ \\
\texttt{$\rho$1e-5}       & 2 &
 $1.15 \times 10^{6} \text{--} 6.3 \times 10^{9}$ & $182$ & 
 $3.0\times10^{9} \text{--} 1.2\times10^{12}$ & $112$ & $-$ & $-$ & $280$ & $48$ \\
\texttt{$\rho$1e-6}       & 2 &
 $1.15 \times 10^{6} \text{--} 6.7 \times 10^{9}$ & $196$ & 
 $3.3\times10^{9} \text{--} 1.7\times10^{12}$ & $126$ & $-$ & $-$ & $308$ & $48$ \\
\texttt{$\rho$1e-7}       & 2 &
 $1.15 \times 10^{6} \text{--} 7.9 \times 10^{9}$ & $196$ & 
 $3.9\times10^{9} \text{--} 2.2\times10^{12}$ & $126$ & $-$ & $-$ & $308$ & $48$ \\
\texttt{$\rho$3e-8}       & 3 &
 $1.15 \times 10^{6} \text{--} 5.2 \times 10^{9}$ & $182$ & 
 $2.5\times10^{9} \text{--} 2.0\times10^{11}$ & $84$ & 
 $4.6\times10^{10} \text{--} 3.7\times10^{12}$ & $84$ & $308$ & $48$ \\
\texttt{$\rho$1e-8}       & 3 &
 $1.15 \times 10^{6} \text{--} 3.8 \times 10^{9}$ & $182$ & 
 $2.0\times10^{9} \text{--} 1.4\times10^{11}$ & $84$ & 
 $5.8\times10^{10} \text{--} 8.0\times10^{12}$ & $98$ & $336$ & $48$ \\
\hline
\end{tabular}
\end{table*}

\subsection{Diagnostics} \label{sec:3.3}

In this section, we define the key quantities used to diagnose the accretion flows: the mass accretion rate, the net force projected along the wind direction, and the rate of energy deposition into the RSG envelope.

The radial mass accretion rate at radius $r$ is given as
\begin{align}
\dot{M}(r)
&= - 2\pi \int_{0}^{\pi} \rho_* v^{r} \, d\theta,\\
&= - 2\pi r^2 \int_{0}^{\pi} \rho u^{r} \sin{\theta} \, d\theta,
\end{align}
where $v^{r}$ is the radial three-velocity.

We pay attention only to the force parallel to the symmetry axis, i.e., the z-component in our coordinates. This is the relevant projected component for orbital evolution in the local wind-tunnel approximation because the relative motion between the NS and the upstream envelope gas is along this axis. We decompose this net projected force into three contributions: $F_\mathrm{mom}$ originated from momentum flux, $F_\mathrm{pre}$ from pressure, and $F_\mathrm{dyn}$ from dynamical gravity. They are defined, respectively, at radius $r$ as
\begin{align}
F_{\mathrm{mom},z}(r)
&= -2\pi \int_{0}^{\pi} S_{i} v^r (e_z)^i d\theta,\\
&= -2\pi r^2 \int_{0}^{\pi} \rho h u_i v^r (e_z)^i \sin{\theta} d\theta,
\end{align}
\begin{align}
F_{\mathrm{pre},z}(r)
&= -2\pi \int_{0}^{\pi} \alpha \sqrt{\gamma} P \delta^r_i (e_z)^i d\theta,\\
&= -2\pi r^2 \int_{0}^{\pi} P \cos{\theta} \sin{\theta} d\theta,
\end{align}
\begin{align}
F_{\mathrm{dyn},z}(r)
= 2\pi
\int_{r}^{R_\mathrm{max}}
\int_{0}^{\pi}
S_0 \partial_i \alpha (e_z)^i d\theta\, dr,
\end{align}
where $R_\mathrm{max}$ denotes the outer radial boundary of the computational domain. The last one is the dynamical gravity exerted by the envelope gas, which is not included in the dynamical equations but is evaluated as a post-processing diagnostic. The total net projected force is the sum of the three contributions:
\begin{equation}
F_{\mathrm{net},z}(r) \equiv F_{\mathrm{mom},z}(r) + F_{\mathrm{pre},z}(r) + F_{\mathrm{dyn},z}(r).
\end{equation}
Throughout this work, we evaluate this force at the NS surface:
\begin{equation}
F_{\mathrm{net}} \equiv F_{\mathrm{net},z}(R_\mathrm{NS}).
\label{eq:Fnet_def}
\end{equation}
We refer to $F_{\rm net}$ as the net projected force. With our sign convention, $F_{\rm net}<0$ corresponds to drag, i.e. a decelerating force opposite to the NS motion, whereas $F_{\rm net}>0$ means thrust, i.e. an accelerating force along the NS motion. These terms are commonly used in the literature \citep[e.g.][]{2017ApJ...838...56M,2019MNRAS.490.3727C}.

Finally, we define the net energy flux through the spherical surface of radius $r$ as
\begin{align}
\dot{E}(r)
&= 2\pi \int_{0}^{\pi} (S_{0}-\rho_*c^2 + \sqrt{\gamma}P) v^r  \, d\theta,\\
&= 2\pi r^2 \int_{0}^{\pi} (\rho h w - \rho c^2) u^r \sin{\theta} \, d\theta.
\end{align}
We are particularly interested in the value at the outer boundary,
$\dot{E}_{\rm out} \equiv \dot{E}(R_\mathrm{max})$,
which represents the energy deposition rate into the RSG envelope.

These diagnostics are time-dependent. We use their values in the (quasi-)steady states alone in Section~\ref{sec:4}.
%-----------------------------------------------------------------
\begin{figure*}
    \centering
    \includegraphics[width=\linewidth]{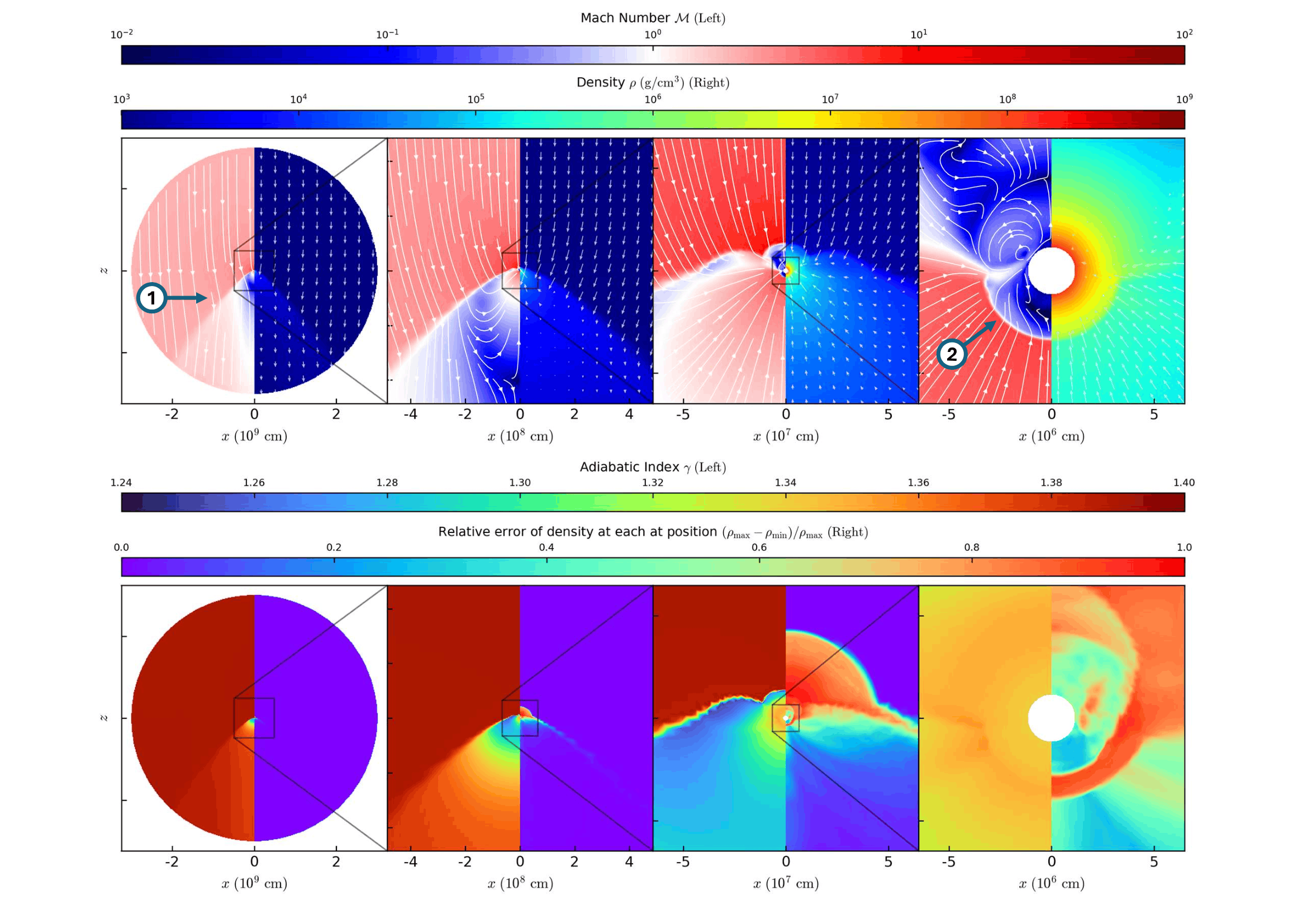}
    \caption{Quasi-steady structure for model \texttt{test-Heos}. Top row: Mach number and streamlines (left half) and density and velocity vectors (right half). Bottom row: adiabatic index (left half) and the time-variability diagnostic defined in Section~\ref{sec:4.1} (right half). Panels from left to right zoom into the inner region by roughly a factor of ten, so that the shock fronts and time variability can be seen clearly. Numbered arrows indicate distinct shocks: (1) the outer bow shock and (2) the inner, oppositely oriented bow shock.}
    \label{fig:test_heos}
\end{figure*}

\section{Results} \label{sec:4}

This section presents the accretion flows we obtain with the multi-layer domain-decomposition method. The domain from the NS surface to the accretion radius is well-resolved everywhere and the accretion flows in the individual layers are combined consistently to produce a single (quasi-)steady flow in the entire computational domain. 

As shown shortly, the flow structures we obtain in our simulations are quite different from those in the classical BHL picture. In Section~\ref{sec:4.1}, we use the test models to demonstrate it. In Section~\ref{sec:4.2}, we show that the same flow structure appears for the more realistic envelope models. Finally, in Section~\ref{sec:4.3}, we evaluate the accretion rates, net projected forces, and energy deposition rate for those flows.

%-------------------------------------------------
\subsection{Flow Morphology: formation of nested shocks and its dependence on EOS} \label{sec:4.1}

We first consider the test models. They are not meant to be very realistic. Instead, they provide a controlled setting, in which the origin of the nested-shock formation, one of our main findings in this paper, can be identified without complications of the multi-layer domain-decomposition. The consistency between the single- and multi-layer computations is validated for the same models in Appendix~\ref{sec:a}.

\begin{figure*}
    \centering
    \includegraphics[width=\linewidth]{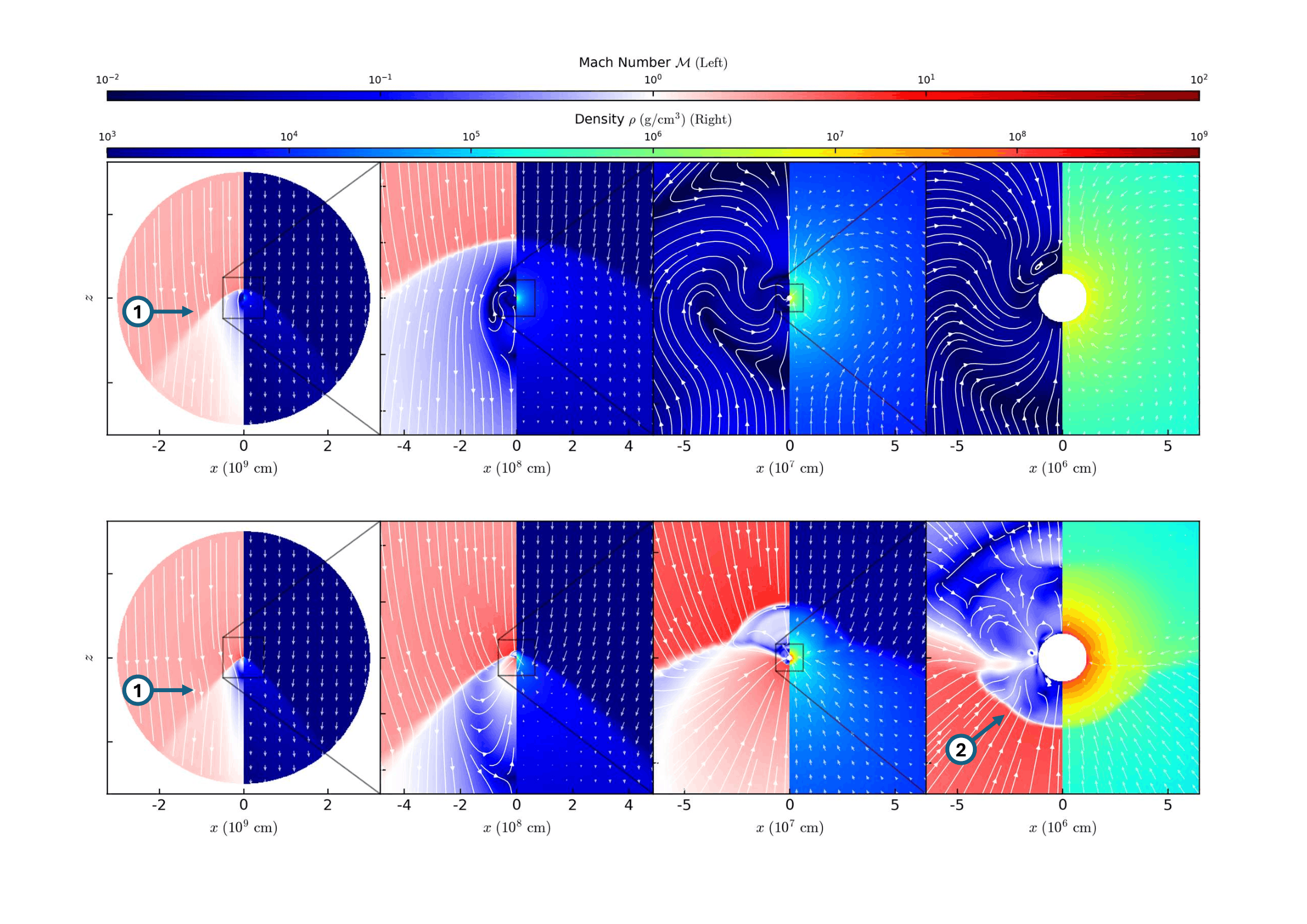}
    \caption{Quasi-steady structures in models \texttt{test-gamma5/3} (top row) and \texttt{test-gamma4/3} (bottom row). For each row, the left half of each panel displays the Mach number with streamlines, and the right half shows the density with velocity vectors as in Fig.~\ref{fig:test_heos}. Numbered arrows follow the same convention as in Fig.~\ref{fig:test_heos}: (1) outer bow shock and (2) inner, oppositely oriented bow shock. In the stiffer EOS case ($\gamma=5/3$), only shock (1) appears and the post-shock flow forms a nearly spherical, quasi-static atmosphere around the NS. In the more compressible case ($\gamma=4/3$), stronger gravitational re-acceleration produces shock (2), yielding a nested-shock morphology.}
    \label{fig:test_gamma}
\end{figure*}

Figure~\ref{fig:test_heos} exhibits the typical nested-shock structure in the quasi-steady accretion flow onto the NS exemplified in the \texttt{test-Heos} run. The panels zoom in on the centre of the computational domain by a factor of 10 from left to right. Note that this model is computed with a single layer. The Mach number map (the left half of each panel on the top row) shows that the incoming supersonic flow first hits an outer bow shock, becoming subsonic. As the shocked gas is gravitationally focused towards the NS, it is re-accelerated; since the flow is eventually brought to rest on the NS surface, the re-accelerated stream must be decelerated again by an inner, oppositely oriented bow shock (referred to as the opposite bow shock). This is the nested-shock structure we find commonly in our models hereafter. The numbered arrows in Fig.~\ref{fig:test_heos} mark these two shocks explicitly: (1) the outer bow shock and (2) the inner, oppositely oriented bow shock.

The compressibility of the Helmholtz EOS is characterized by the adiabatic index as $\gamma \equiv(\partial\ln P/\partial\ln\rho)_{s}$; unlike in the gamma-law EOS, in which $\gamma$ is constant, it varies with $(\rho,T)$ (and composition). The value of $\gamma$ therefore changes with radius in the RSG envelope; it can also change across shocks. In particular, close to the NS we find $\gamma<4/3$, associated with electron--positron pair creation at high temperatures. The resultant enhancement of compressibility makes it easier for the post-shock flow to re-accelerate to supersonic speeds, thereby enabling the formation of the inner shock.

The time-variability map (the right half of each panel on the bottom row in Fig.~\ref{fig:test_heos}) presents a measure of steadiness. We define the normalized density fluctuation as $(\rho_{\max}-\rho_{\min})/\rho_{\max}$, where $\rho_{\max}$ and $\rho_{\min}$ are the maximum and minimum densities attained at each spatial location over a duration of $\sim 10$ dynamical times in the quasi-steady state. Values near zero of this quantity imply an approximately steady state, whereas values close to unity indicate strong time variations. Although both shock waves continue to oscillate around their average positions in this model---probably consistent with shock oscillations driven by the entropic-acoustic instability \citep{2005A&A...435..397F}---the variability is small throughout most of the computational domain. We therefore refer to the overall configuration thus obtained as ``quasi-steady.''

Figure~\ref{fig:test_gamma} compares the results of test runs (models \texttt{test-$\gamma$5/3} and \texttt{test-$\gamma$4/3}) with the gamma-law EOS: $\gamma=5/3$ (top) and $\gamma=4/3$ (bottom). For the stiffer case of $\gamma=5/3$, the post-shock gas remains subsonic down to the NS, forming a nearly hydrostatic and spherical atmosphere around it; as a result, no additional inner shock develops in this case.

In contrast, for the more compressible case of $\gamma=4/3$, the shocked gas can be focused more easily and re-accelerated more strongly as it falls deeper in the NS potential. As in model \texttt{test-Heos}, the re-accelerated supersonic stream is re-decelerated by an inner, oppositely oriented shock, yielding a nested-shock morphology. It is noted that unlike model test-Heos with the Helmholtz EOS, this model with $\gamma=4/3$ does not have the region in the vicinity of NS, where the re-acceleration is enhanced by copious electron--positron pair creation.

%-------------------------------------------------
\subsection{Dependence on accretion radius in realistic envelope conditions} \label{sec:4.2}

We next examine models whose outer boundary conditions $(\rho_\infty, T_\infty, v_\infty)$ are adopted from the realistic RSG envelope profiles calculated by MESA (Table~\ref{tab:twocol1}). These runs probe how the shock morphology identified in the test models manifests under the realistic conditions and how it varies with the accretion radius $R_a$. The flows are computed with two or three layers depending on $R_a$ (see Table~\ref{tab:layers}). 

\begin{figure*}
    \centering
    \includegraphics[width=\linewidth]{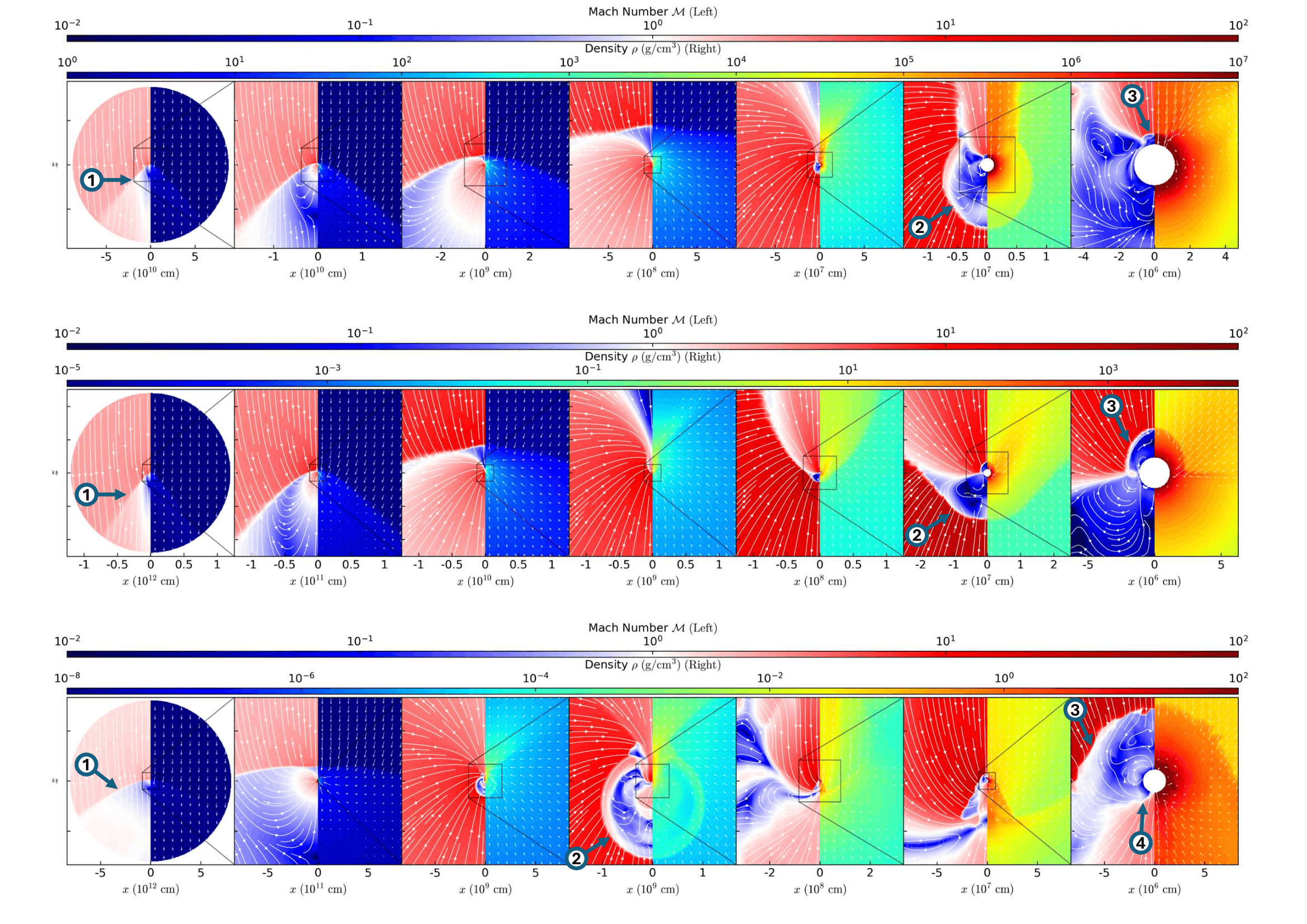}
    \caption{Representative realistic models corresponding to different locations in the RSG envelope. Top, middle, and bottom rows correspond to $\rho_\infty=1$, $10^{-5}$, and $10^{-8}\,\mathrm{g\,cm^{-3}}$. As in Fig.~\ref{fig:test_gamma}, the left and right halves of each panel display the Mach number with streamlines and the density with velocity vectors, respectively. Panels from left to right zoom into the inner region by roughly a factor of ten, so that the shock fronts can be seen clearly. Numbered arrows mark the positions of distinct shock waves.}
    \label{fig:rho1d_combined}
\end{figure*}

\begin{figure}
    \centering
    \includegraphics[width=\linewidth]{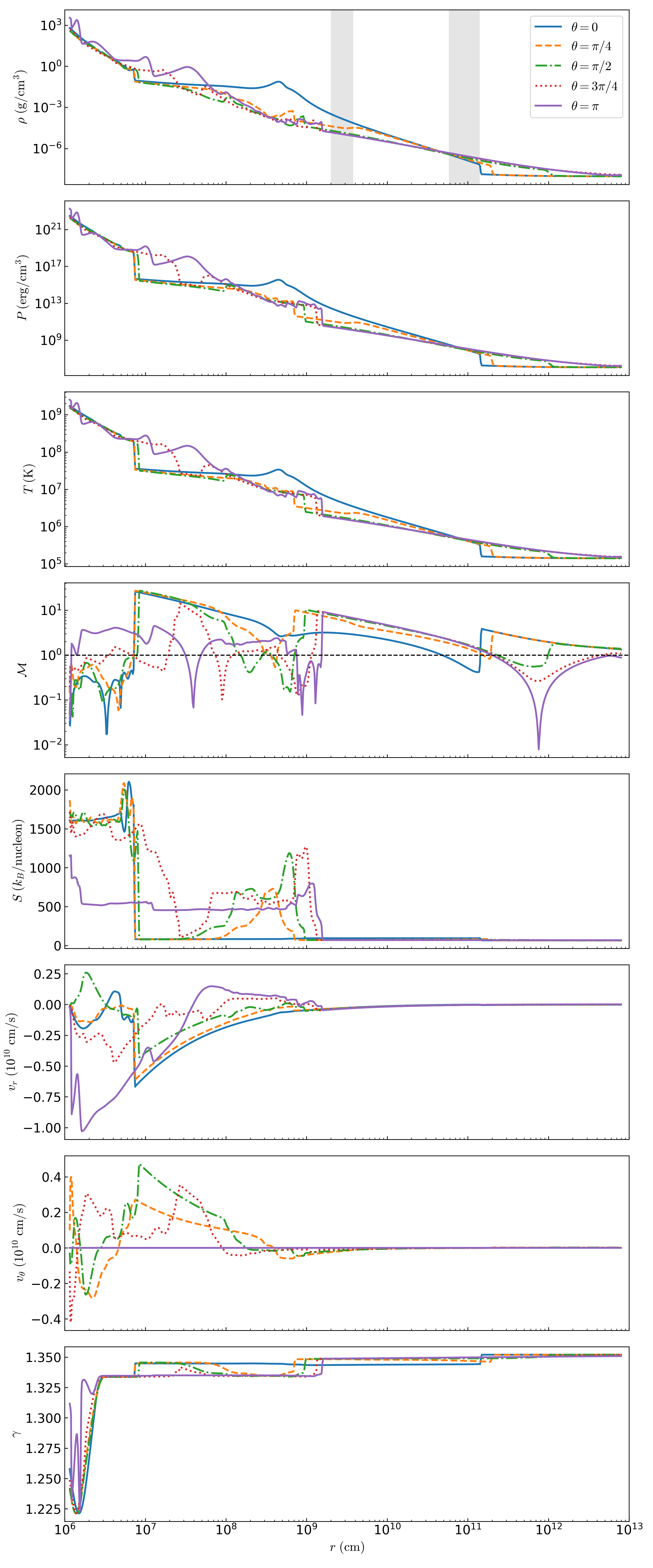}
    \caption{Radial profiles from the NS surface to the outer boundary for model \texttt{$\rho$1e-8} with $\rho_\infty=10^{-8}\mathrm{g\,cm^{-3}}$. Shown from top to bottom are density, temperature, Mach number, entropy per nucleon, radial and $\theta$ components of velocity, and adiabatic index. Different lines denote 5 directions from $\theta=0$ to $\theta=\pi$ in increments of $\pi/4$. The two grey shaded areas in the density panel indicate the overlapping regions resulting from the three-layer domain-decomposition.}
    \label{fig:rho1d_minus8_radial_profile}
\end{figure}

Figure~\ref{fig:rho1d_combined} shows representative accretion flows for different $R_a$, plotting the Mach number and streamlines on the left half and the density and velocity vectors on the right half just as in Figs.~\ref{fig:test_heos} and~\ref{fig:test_gamma}. Each row of panels progressively zooms into the inner region by roughly a factor of ten, and numbered arrows indicate the positions of distinct shock waves.

In the top row ($\rho_\infty=1\,\mathrm{g\,cm^{-3}}$), three shocks are identified: in addition to the outer bow shock and the opposite bow shock found in the test models, a third shock wave emerges near the surface of the NS. The gas falls supersonically towards the NS from almost all directions at radii of a few hundred kilometres; therefore, the oscillations of the innermost shock do not propagate upstream to affect the outer two bow shocks.

A similar decoupling is observed in the intermediate case shown in the middle row ($\rho_\infty=10^{-5}\,\mathrm{g\,cm^{-3}}$), where the numbered arrows again mark three shocks. Furthermore, there is a clear trend that as the accretion radius $R_a$ increases, the primary shock front expands outward (see the scale difference between the two panels). Since $R_a$ is approximately 400 times larger in this model than that in the test model, the shortest distance from the NS to the shock front (standoff distance) also increases by roughly a factor of 20 to $4 \times 10^9\,\mathrm{cm}$.

As $R_a$ increases further, equivalently for lower $\rho_\infty$ (see the bottom row), the shocked region becomes more extended, allowing the flow to undergo additional sonic transitions and develop more complex shock morphologies. In the bottom row ($\rho_\infty=10^{-8}\,\mathrm{g\,cm^{-3}}$), four shock fronts are identified by the numbered arrows. The outermost shock is steady, whereas the second outermost shock oscillates over a spatial scale of order $10^9\,\mathrm{cm}$, and the innermost two shocks confined by the outer two shocks are also oscillatory within a region of order $10^7\,\mathrm{cm}$. Thus, only the outermost shock is (almost) steady. Note that it is quite distant from the other shocks.

Figure~\ref{fig:rho1d_minus8_radial_profile} complements the 2D morphologies presented in Fig.~\ref{fig:rho1d_combined} by illustrating how key thermodynamic and kinematic quantities vary with radius. As detailed in the caption, the profiles are shown for 5 different directions ($\theta=0$ to $\pi$ in steps of $\pi/4$), and the grey shaded areas in the density panel mark the overlapping regions of the three computational layers. From these plots, we can clearly trace the flow dynamics: across the shock, the flow transitions from supersonic to subsonic, accompanied by a sharp rise in density, temperature, and entropy per nucleon. The Mach number and radial velocity profiles demonstrate that the flow is re-accelerated towards the NS downstream of the shock. This renewed gravitational acceleration leads to the formation of an inner shock transition shown in the 2D maps. The $\theta$ component of velocity becomes non-negligible in the shocked region, reflecting the emergence of circulation generated by the bow-shock geometry. As noted for the test models, the adiabatic index drops below $4/3$ at $r \lesssim 10^9\,\mathrm{cm}$ due to electron--positron pair creation, which is also vindicated by the rise of entropy per baryon to $\gtrsim 500\,\mathrm{k_B}$.

%-------------------------------------------------
\subsection{Evaluation of diagnostic quantities: mass accretion rate, net projected force, and energy deposition rate} \label{sec:4.3}

We now quantify how the nested-shock structure affects the mass accretion rate, net projected force, and energy deposition rate. Table~\ref{tab:results_summary} summarizes the quasi-steady values measured in our models; parenthetical numbers denote the ratio to the corresponding BHL momentum-accretion estimates ($\dot{M}/\dot{M}_{\rm BHL}$, $F_{\rm net}/F_{\rm BHL}$, and $\dot{E}_{\rm out}/\dot{E}_{\rm BHL}$, respectively). Here $F_{\rm BHL}$ is the positive reference magnitude defined by Eq.~(\ref{eq:f_bhl}). We recall our sign convention (Section~\ref{sec:3.3}): $F_{\rm net}<0$ corresponds to drag, whereas $F_{\rm net}>0$ means thrust.

\begin{figure}
    \centering
    \includegraphics[width=\linewidth]{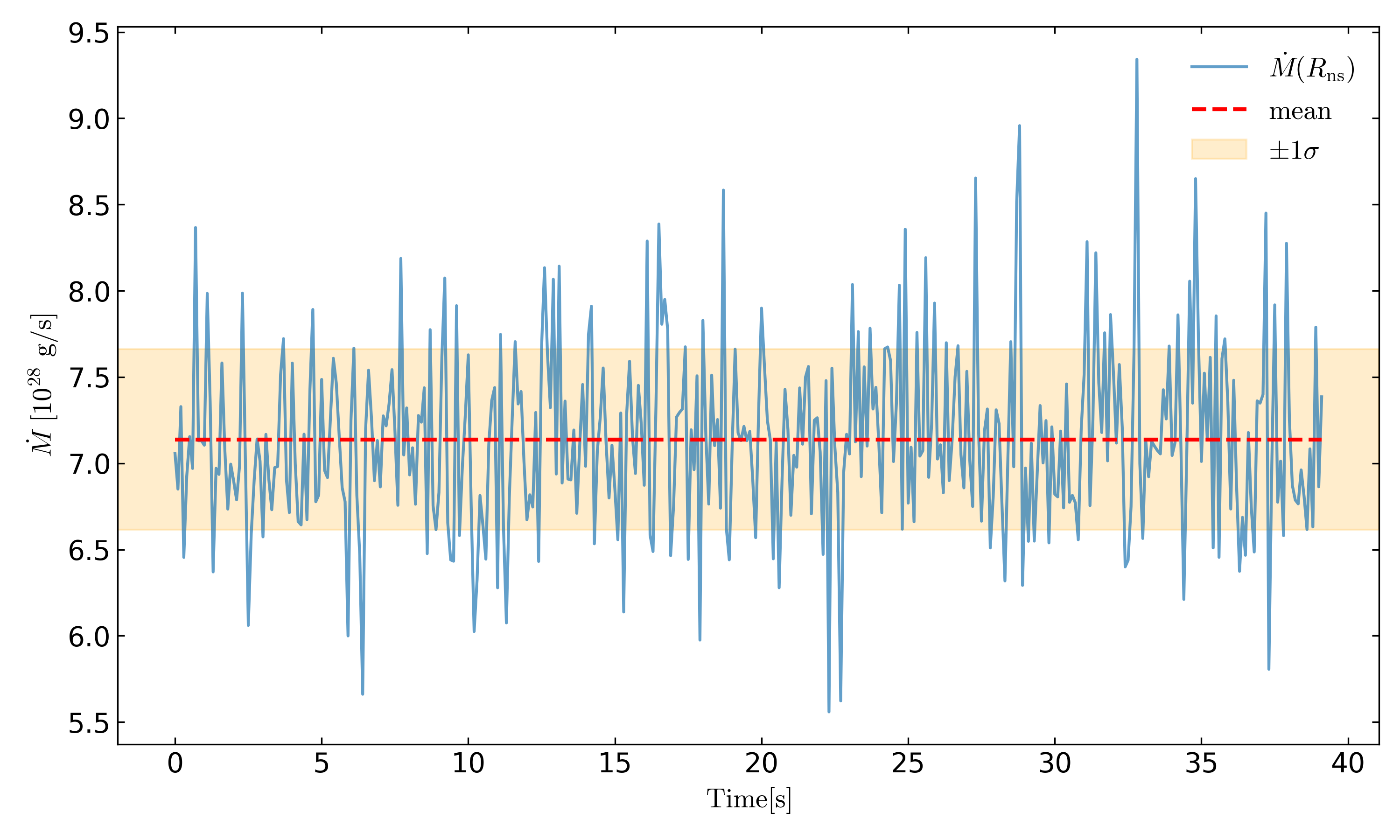}
    \caption{Time evolution of the mass accretion rate $\dot{M}(R_{\rm NS})$ at the NS surface for model \texttt{$\rho$1e0} with $\rho_\infty=1\,\mathrm{g\,cm^{-3}}$. The blue curve shows the instantaneous value; the red dashed line indicates the time-averaged value; the orange shaded band corresponds to the $\pm 1\sigma$ range.}
    \label{fig:rho1d_0_mdot}
\end{figure}

\begin{figure}
    \centering
    \includegraphics[width=\linewidth]{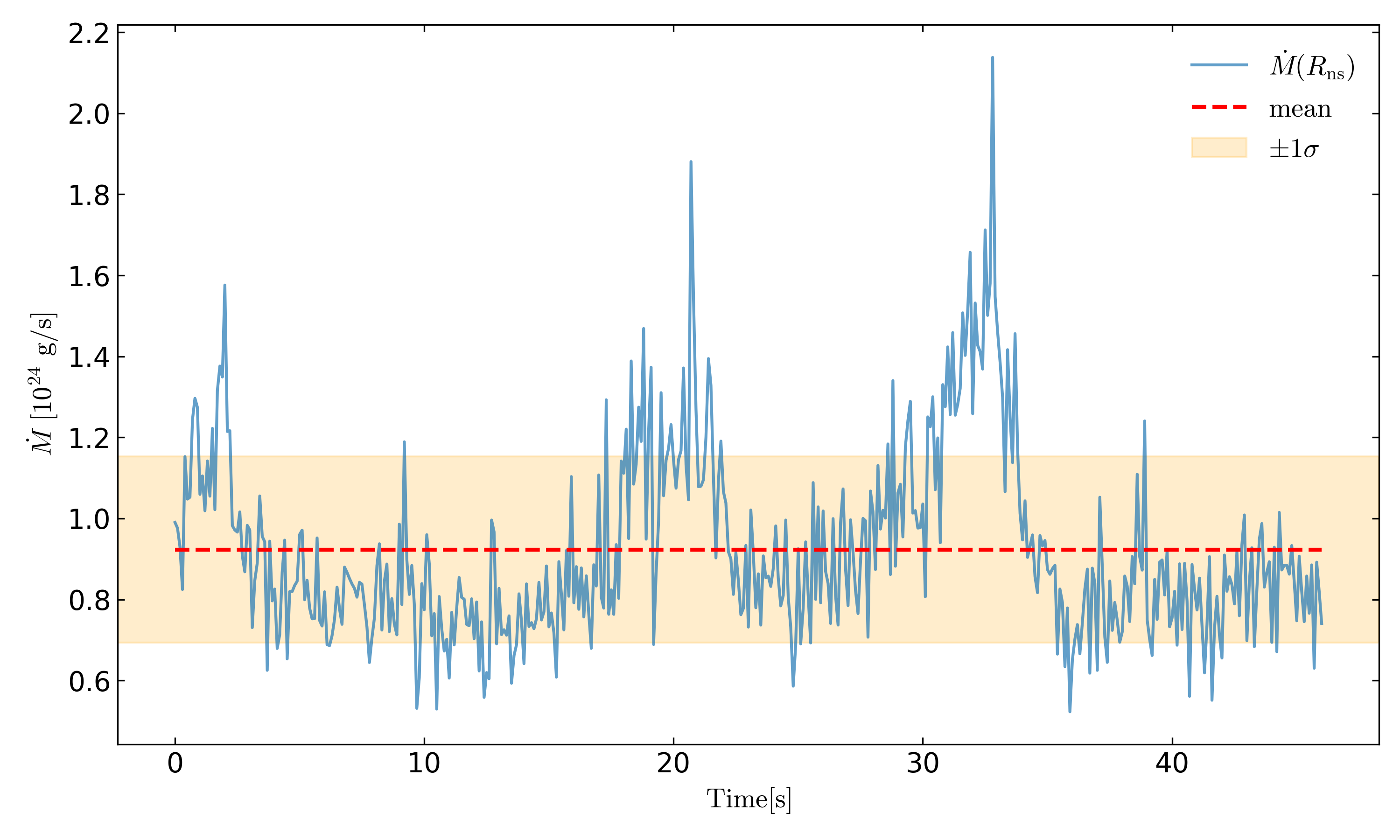}
    \caption{Same as Fig.~\ref{fig:rho1d_0_mdot} but for model \texttt{$\rho$1e-8} with $\rho_\infty=10^{-8}\,\mathrm{g\,cm^{-3}}$. Note larger fluctuations compared with the high-density case.}
    \label{fig:rho1d_minus8_mdot}
\end{figure}

Figures~\ref{fig:rho1d_0_mdot} and \ref{fig:rho1d_minus8_mdot} show the time evolution of the mass accretion rate $\dot{M}(R_{\rm NS})$ measured at the NS surface for representative high-density ($\rho_\infty=1\,\mathrm{g\,cm^{-3}}$) and low-density ($\rho_\infty=10^{-8}\,\mathrm{g\,cm^{-3}}$) models. In both cases, $\dot{M}$ does not settle to a constant value but fluctuates persistently, reflecting the ongoing shock oscillations near the NS surface noted in Section~\ref{sec:4.1}. Nevertheless, the temporal mean (red dashed line) is well defined with fluctuation amplitude of $~7\%$ and $~20\%$ at $1\sigma$, respectively, allowing us to extract a meaningful (in the time-averaged sense) accretion rate.

The character of the time variability, however, differs between the two models. In the high-density case (Fig.~\ref{fig:rho1d_0_mdot}), the fluctuations are relatively regular and of moderate amplitude: even the peak-to-peak variation is roughly $\pm 15$--$20\%$ of the mean, consistent with the quasi-periodic shock oscillations confined near the NS surface (see Section~\ref{sec:4.1}). In the low-density case (Fig.~\ref{fig:rho1d_minus8_mdot}), by contrast, the fluctuations are larger and more intermittent, with sporadic bursts reaching up to twice the mean value. This enhanced variability reflects the more complex and spatially extended shock structure present in large-$R_a$ models (Section~\ref{sec:4.2}): since the (almost steady) outermost shock is farther apart from the other shocks in these models, the oscillations of the second outer shock span a wider radial extent interior to the outermost shock (compared with the high-density models $\texttt{$\rho$1e0}$ and $\texttt{$\rho$1e-5}$) and drive larger excursions in the instantaneous accretion rate as a result.

Despite these differences in variability, the time-averaged $\dot{M}$ remains within a factor of about 2 of the BHL estimate across the entire set of models surveyed, with $\dot{M}/\dot{M}_{\rm BHL}\simeq 1.3$--$2.1$ (Table~\ref{tab:results_summary}). This indicates that the net mass supply is governed primarily by the gravitational focusing on the accretion-radius scale and by the outermost bow shock, and is rather insensitive to the details of the inner-shock dynamics.

\begin{table*}
\centering
    \caption{Summary of quasi-steady accretion rates, net projected forces, and energy deposition rates at the outer boundary across the model survey.}
\label{tab:results_summary}
\begin{tabular}{ccccccc}
\hline
model &
$\dot{M}$ (g/s) &
$F_{\mathrm{mom},z}$ (dyn) &
$F_{\mathrm{pre},z}$ (dyn) &
$F_{\mathrm{dyn},z}$ (dyn) &
$F_{\mathrm{net}}$ (dyn) &
$\dot{E}_{\rm out}$ (erg/s) \\
\hline
\texttt{test-Heos} &
$5.0 \times 10^{29} (1.6)$ &
$-5.4 \times 10^{37} $& 
$-2.3 \times 10^{39} $&
$ 3.0 \times 10^{38} $&
$-2.1 \times 10^{39} (-6.5)$ &
$5.1 \times 10^{47} (1.6)$ \\
\texttt{test-$\gamma$5/3} &
$1.8 \times 10^{28} (0.056)$ &
$2.4 \times 10^{35} $&
$-3.7 \times 10^{36} $&
$-2.4 \times 10^{39} $&
$-2.4 \times 10^{39} (-7.8)$ &
$1.2 \times 10^{46} (0.038)$ \\
\texttt{test-$\gamma$4/3} &
$4.5 \times 10^{29} (1.4)$ &
$-3.7 \times 10^{37} $&
$-1.4 \times 10^{39} $&
$-6.1 \times 10^{38} $&
$-2.0 \times 10^{39} (-6.4)$ & 
$5.7 \times 10^{47} (1.8)$ \\
\texttt{$\rho$1e0} &
$ 7.1 \times 10^{28} (1.4)$ &
$-3.9 \times 10^{37} $&
$-9.5 \times 10^{38} $&
$ 1.2 \times 10^{39} $&
$ 2.4 \times 10^{38} (12)$ &
$ 2.7 \times 10^{45} (1.6)$ \\
\texttt{$\rho$1e-1} &
$ 1.5 \times 10^{28} (1.5)$ &
$-9.0 \times 10^{36} $&
$-2.5 \times 10^{38} $&
$ 3.1 \times 10^{38} $&
$ 5.4 \times 10^{37} (15)$ &
$ 3.6 \times 10^{44} (1.7)$ \\
\texttt{$\rho$1e-2} &
$ 3.4 \times 10^{27} (1.4) $ &
$ 8.8 \times 10^{34} $&
$-6.1 \times 10^{37} $&
$ 5.8 \times 10^{37} $&
$-3.4 \times 10^{36} (-13)$ &
$ 4.5 \times 10^{43} (1.6)$ \\
\texttt{$\rho$1e-3} &
$ 7.1 \times 10^{26} (1.4)$ &
$-3.3 \times 10^{34} $&
$-7.5 \times 10^{36} $&
$ 8.5 \times 10^{36} $&
$ 9.3 \times 10^{35} (21)$  &
$ 5.5 \times 10^{42} (1.5)$  \\
\texttt{$\rho$1e-4} &
$1.4 \times 10^{26} (1.4)$ &
$-3.3 \times 10^{33} $&
$-7.5 \times 10^{35} $&
$ 9.8 \times 10^{35} $&
$ 2.3 \times 10^{35} (33)$ &
$ 7.7 \times 10^{41} (1.7)$ \\
\texttt{$\rho$1e-5} &
$ 3.1 \times 10^{25} (1.3)$ &
$-1.2 \times 10^{31} $&
$ 3.3 \times 10^{35} $&
$-3.2 \times 10^{35} $&
$ 1.5 \times 10^{34} (12)$ &
$ 8.6 \times 10^{40} (1.3)$ \\
\texttt{$\rho$1e-6} &
$ 5.5 \times 10^{24} (1.3)$ &
$-2.7 \times 10^{31} $&
$ 1.0 \times 10^{35} $&
$-9.4 \times 10^{34} $&
$ 7.6 \times 10^{33} (42)$ & 
$ 1.1 \times 10^{40}$ (1.4) \\
\texttt{$\rho$1e-7} &
$ 1.1 \times 10^{24} (1.8)$ &
$ 2.1 \times 10^{31} $&
$ 1.6 \times 10^{34} $&
$-1.4 \times 10^{34} $&
$ 2.1 \times 10^{33} (93)$ &
$ 2.2 \times 10^{39}$ (2.9) \\
\texttt{$\rho$3e-8} &
$ 8.8 \times 10^{23} (2.1)$ &
$ 4.8 \times 10^{32} $ &
$ 1.2 \times 10^{34} $ &
$-1.4 \times 10^{34} $ &
$-1.2 \times 10^{33} (-114)$ &
$ 1.2 \times 10^{39}$ (4.2) \\
\texttt{$\rho$1e-8} &
$ 9.0 \times 10^{23} (2.1)$ &
$ 2.1 \times 10^{31}$ &
$ 6.9 \times 10^{33}$ &
$-7.7 \times 10^{33}$ &
$-7.9 \times 10^{32} (-106)$ &
$ 6.0 \times 10^{38}$ (4.7) \\
\hline
\end{tabular}
\\
\raggedright
\textit{Note.} The numbers in parentheses show the values scaled by the corresponding BHL momentum-accretion estimates (Eqs.~\ref{eq:mdot_bhl}--\ref{eq:edot_bhl}) derived in Section~\ref{sec:2.1}.
\end{table*}

\begin{figure}
    \centering
    \includegraphics[width=\linewidth]{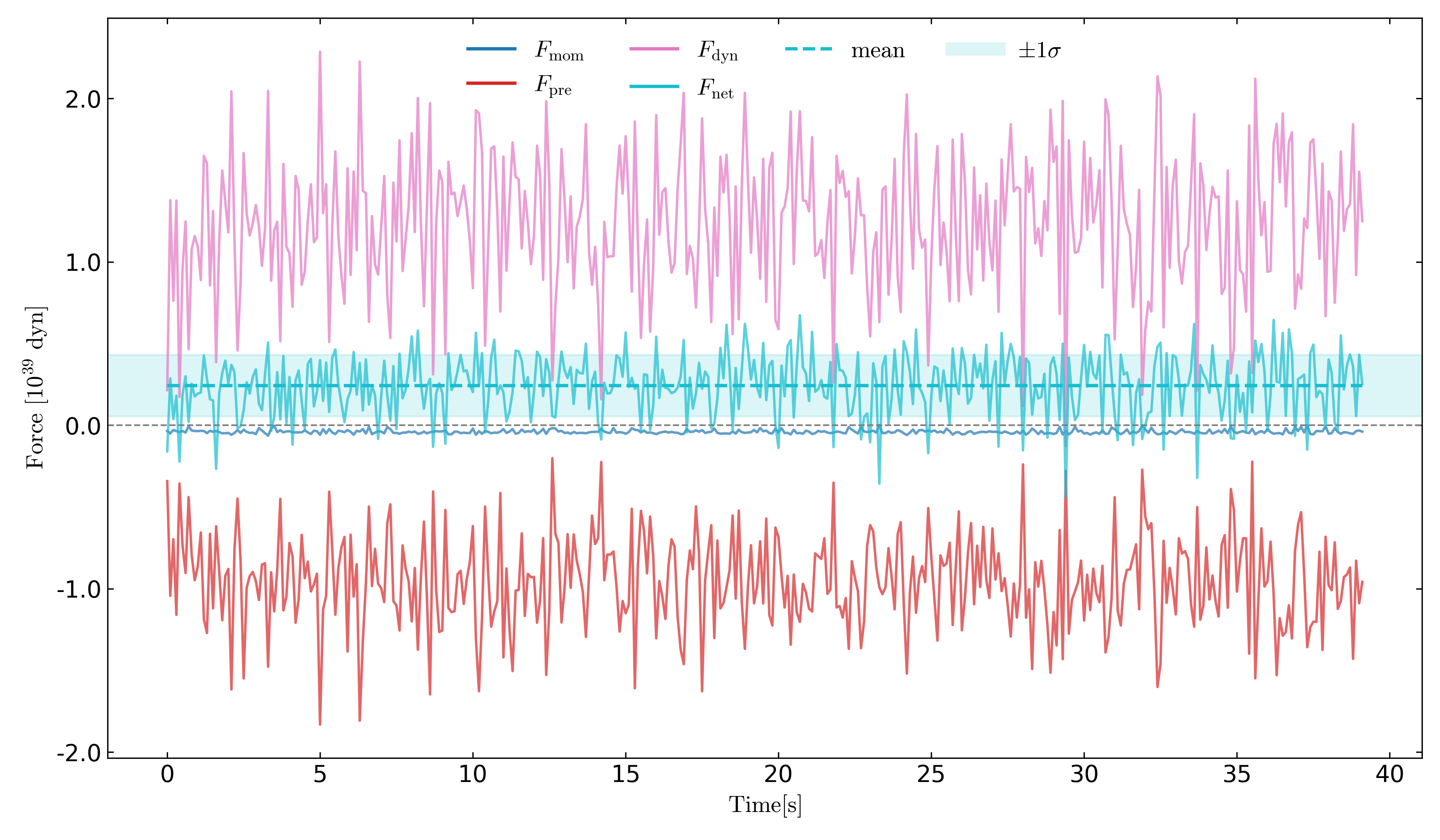}
    \caption{Time evolution of the net projected force at the NS surface for model \texttt{$\rho$1e0} with $\rho_\infty=1\,\mathrm{g\,cm^{-3}}$. Colours denote different contributions (\ref{eq:Fnet_def} in Sec.~\ref{sec:3.3}): $F_{\mathrm{mom},z}$ (blue), $F_{\mathrm{pre},z}$ (red), $F_{\mathrm{dyn},z}$ (magenta), and the net force $F_{\mathrm{net}}$ (cyan). The cyan dashed line and shaded band show the time-averaged value and $\pm 1\sigma$ dispersion of $F_{\mathrm{net}}$, respectively.}
    \label{fig:rho1d_0_drag}
\end{figure}

\begin{figure}
    \centering
    \includegraphics[width=\linewidth]{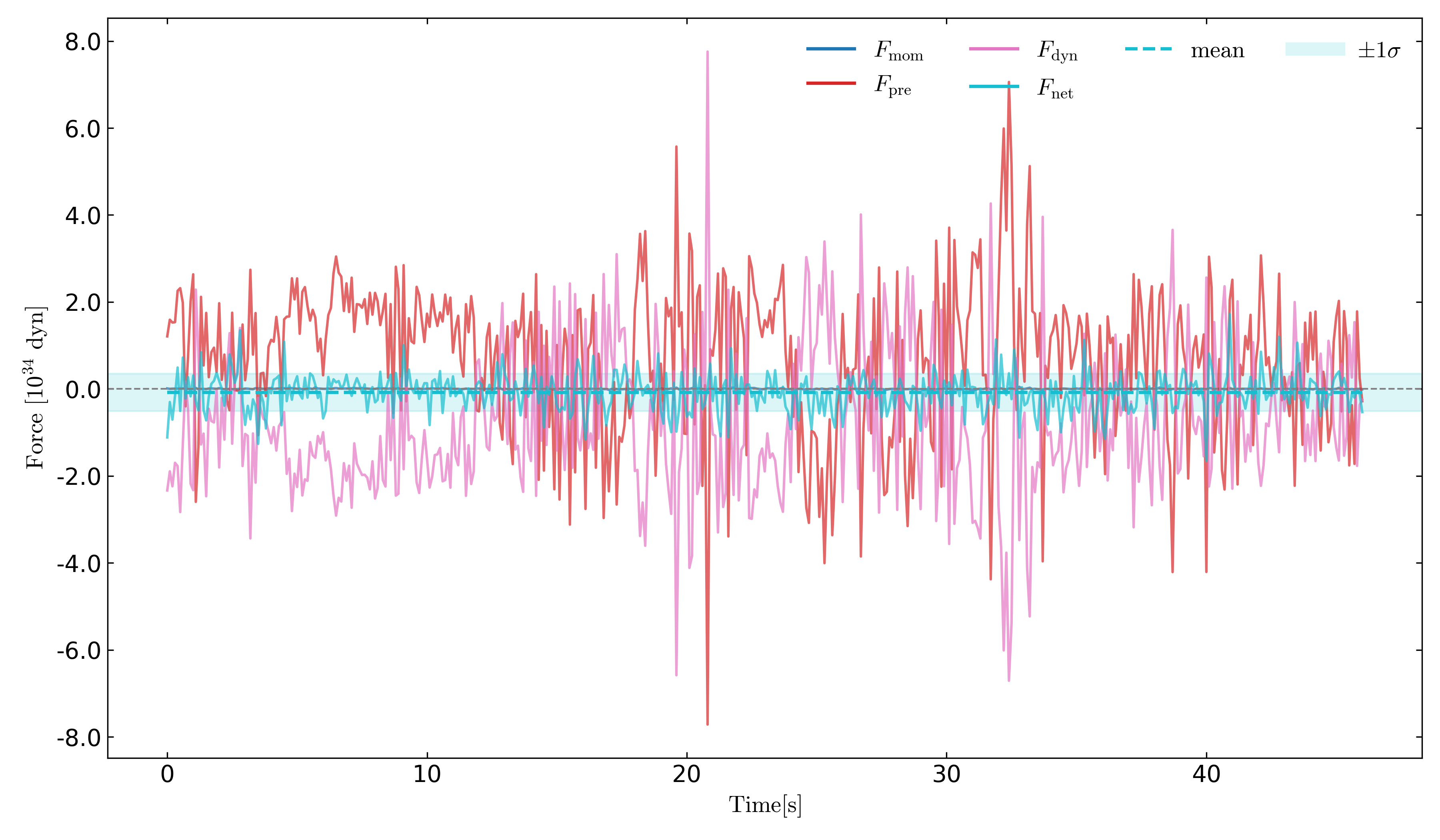}
    \caption{Same as Fig.~\ref{fig:rho1d_0_drag} but for model \texttt{$\rho$1e-8} with $\rho_\infty=10^{-8}\,\mathrm{g\,cm^{-3}}$. $F_{\mathrm{pre},z}$ (red) and $F_{\mathrm{dyn},z}$ exhibit much larger fluctuations than $F_{\mathrm{net}}$, essentially the sum thereof.}
    \label{fig:rho1d_minus8_drag}
\end{figure}

The net projected force $F_{\rm net}$ (cyan) is decomposed into three contributions (see Sec.~\ref{sec:3.3}): the momentum-flux term $F_{\mathrm{mom},z}$, the pressure term $F_{\mathrm{pre},z}$, and the dynamical-gravity term $F_{\mathrm{dyn},z}$. Figures~\ref{fig:rho1d_0_drag} and \ref{fig:rho1d_minus8_drag} show their time evolutions at the NS surface for the same representative models discussed in the previous section.

A common feature across both models is that $F_{\mathrm{mom},z}$ is negligible compared with the other two components, $F_{\mathrm{pre},z}$ and $F_{\mathrm{dyn},z}$, which are opposite in sign. In the high-density case (Fig.~\ref{fig:rho1d_0_drag}), $F_{\mathrm{dyn},z}$ fluctuates around ${\sim}\,+1.2\times 10^{39}\,\mathrm{dyn}$ and $F_{\mathrm{pre},z}$ around ${\sim}\,-9.5\times 10^{38}\,\mathrm{dyn}$; their partial cancellation yields a net positive force $F_{\rm net}\sim +2.4\times 10^{38}\,\mathrm{dyn}$ on average. With the sign convention defined in Section~\ref{sec:3.3}, this is not a drag but a thrust. Note also that the fluctuations of $F_{\rm net}$ are moderate and the time-averaged value (dashed line) is well-defined and meaningful.

In the low-density case (Fig.~\ref{fig:rho1d_minus8_drag}), the cancellation between $F_{\mathrm{pre},z}$ and $F_{\mathrm{dyn},z}$ is far more severe: both components individually reach amplitudes of several $\times 10^{34}\,\mathrm{dyn}$ with violent, intermittent fluctuations, yet the net projected force remains small after cancellation, with a time-averaged value of $-7.9\times10^{32}\,\mathrm{dyn}$ and a $1\sigma$ dispersion of $4.4\times10^{33}\,\mathrm{dyn}$. This demonstrates that $F_{\rm net}$ is an outcome of the subtle instantaneous imbalance between the local pressure and dynamical gravity in the wake, which are strongly (anti-)correlated with each other.

In contrast to $\dot{M}$, which remains within a factor of about 2 of the BHL estimate, the time-averaged $|F_{\rm net}|$ departs from the BHL momentum-accretion estimate by one to two orders of magnitude (Table~\ref{tab:results_summary}). This demonstrates that such broad agreement in $\dot{M}$ does not imply that the BHL momentum-accretion model is a good approximation: whereas the large-scale inflow is set by gravitational focusing on the accretion scale as the BHL model shows, the net projected force is determined by the delicate force balance in the inner region, where non-spherical pressure and the dynamical gravity from shocked matter in the wake dominate. As $R_a$ increases, these contributions become more important, and the ratio $|F_{\rm net}|/F_{\rm BHL}$ deviates rapidly from unity even when $\dot{M}/\dot{M}_{\rm BHL}$ remains of order unity.

Table~\ref{tab:results_summary} also shows that the sign of $F_{\rm net}$ is not necessarily negative as expected intuitively. Among the realistic models, those with $\rho_\infty = 10^{-2}\,\mathrm{g\,cm^{-3}}$ (\texttt{$\rho$1e-2}), $3\times10^{-8}\,\mathrm{g\,cm^{-3}}$ (\texttt{$\rho$3e-8}), and $10^{-8}\,\mathrm{g\,cm^{-3}}$ (\texttt{$\rho$1e-8}) yield $F_{\rm net}<0$ (drag), while the remaining models obtain $F_{\rm net}>0$ (thrust). This is in sharp contrast to the BHL momentum-accretion model, in which the drag vector always points opposite to the motion, whereas $F_{\rm BHL}$ itself is defined as its positive reference magnitude. The positive $F_{\rm net}$ indicates that the detailed inner-flow structure---in particular, the asymmetric pressure and density distributions produced by the nested shocks---can reverse the direction of the net projected force. It should be remembered, however, that the sign of $F_{\rm net}$ is a result of strong cancellation between $F_{\mathrm{pre},z}$ and $F_{\mathrm{dyn},z}$. As shown by the inner-boundary and resolution tests in Appendices~\ref{sec:d} and \ref{sec:e}, the individual components of the net projected force are not fully converged. In addition, these tests were conducted only for \texttt{test-Heos}, which has $F_{\rm net}<0$, and hence did not establish the numerical robustness of the results for the realistic models, particularly those with $F_{\rm net}>0$. Moreover, the results may change if we incorporate the gravitational backreaction of the wake self-consistently in the simulations. We should therefore regard the occurrence of thrust in some models as tentative. The sign issue aside, the magnitude $|F_{\rm net}|$ is consistently larger than the BHL momentum-accretion estimate by one to two orders of magnitude in all models. We regard this magnitude enhancement as the more robust quantitative result of this study.

Finally, we evaluate the rate of energy deposition into the RSG envelope at the outer boundary, $\dot{E}_{\rm out}$, using the definition in Section~\ref{sec:3.3}. It is the net energy flux going out of the outer boundary of the simulation domain. The resulting values are listed in the rightmost column of Table~\ref{tab:results_summary}. The ratio $\dot{E}_{\rm out}/\dot{E}_{\rm BHL}$ ranges from $\sim 1.3$ to $\sim 4.7$, showing a moderate but systematic enhancement over the BHL estimate. Models with larger $R_a$ tend to have higher energy deposition rates. This is a consequence of the increasingly extended and nested morphology seen in the low-density (and therefore larger $R_a$) models (Section~\ref{sec:4.2}). The infalling gas experiences repeated heating by the nested shocks, efficiently converting its kinetic energy into thermal energy in those models.

To summarize this section, we find that the time-averaged $\dot{M}$ remains within a factor of about 2 of the BHL prediction. This does not imply that the BHL momentum-accretion model is a good surrogate for the realistic simulations. In fact, the energy deposition rate at the outer boundary, $\dot{E}_{\rm out}$, shows a moderate enhancement and, most strikingly, the magnitude of the time-averaged net projected force, $|F_{\rm net}|$, exceeds the BHL momentum-accretion estimate by one to two orders of magnitude after strong cancellation between the local pressure and dynamical gravity, both of which have large magnitudes individually (see Figs.~\ref{fig:rho1d_0_drag} and \ref{fig:rho1d_minus8_drag}). This marked enhancement in the local net projected force could have important implications for orbital evolution during CEE if it persists in a global setting, as discussed next in Section~\ref{sec:5}. The auxiliary tests seem to support this picture: Appendix~\ref{sec:b} shows that relativistic corrections to the force are modest but non-negligible, Appendix~\ref{sec:c} recovers the same nested-shock morphology with Athena++, and Appendix~\ref{sec:e} shows that the enhancement of $|F_{\rm net}|$ over the BHL momentum-accretion estimate persists and strengthens with increasing resolution. Note, however, that the individual components of the net projected force are not fully convergent with the current resolution; the tests were restricted to model \texttt{test-Heos}, in which $F_{\rm net}<0$, and hence did not establish the numerical robustness of the results for the realistic models, particularly those with $F_{\rm net}>0$.

%-----------------------------------------------------------------
\section{Implications for the spiral-in phase} \label{sec:5}
\subsection{Orbital evolution with a simulation-calibrated net-force prescription} \label{sec:5.1}

Our simulations show that the magnitude of the time-averaged net projected force acting on the embedded NS exceeds the canonical BHL momentum-accretion estimate by one to two orders of magnitude (Section~\ref{sec:4.3}). The orbital-energy change during spiral-in is governed by the work done by the net projected force, $\dot{E}_{\rm orb} \simeq F_{\rm net}\, v_{\rm orb}$, in the local approximation and the enhancement in $|F_{\rm net}|$ could substantially modify the spiral-in trajectory if it persisted. To gauge the maximum possible impact, we integrate the following equations to obtain the orbit starting at $a = 0.95\,\mathrm{R_\mathrm{RSG}} \simeq 2300\,R_\odot$:

\begin{figure}
    \centering
    \includegraphics[width=\linewidth]{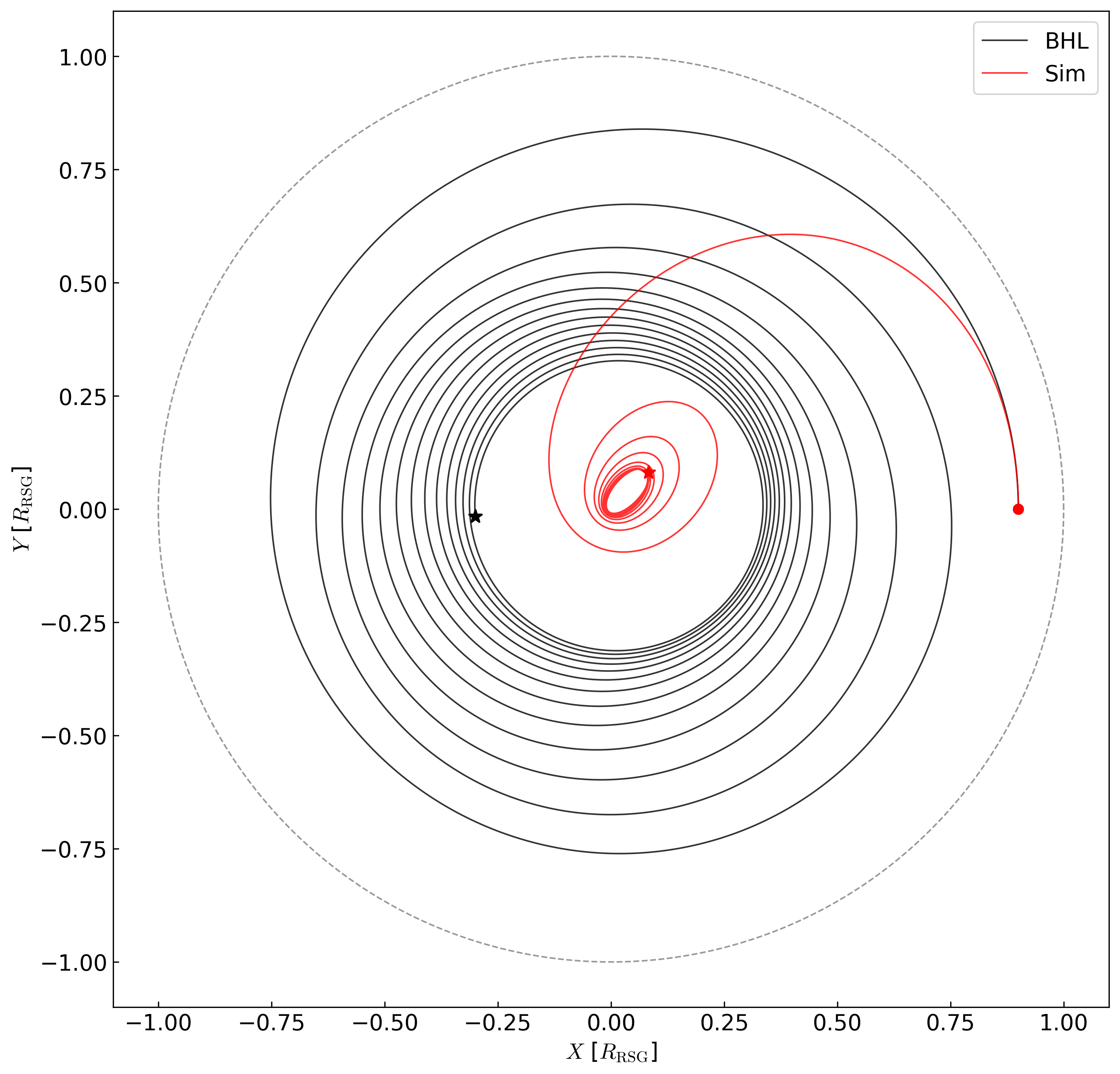}
    \caption{Spiral-in trajectories of an NS inside an RSG envelope, starting from the same initial condition at $a = 0.95\,\mathrm{R_\mathrm{RSG}} \simeq 2300\,R_\odot$. The black curve is obtained by a numerical integration of Eqs.~(\ref{eq:orbit_r})--(\ref{eq:orbit_vphi}) with the canonical BHL momentum-accretion prescription for 40\,yr; the red curve shows the corresponding trajectory for 4\,yr under the simulation-calibrated prescription in which the net projected force inferred in this work is applied. The filled circle and star correspond to the initial and final positions, respectively.}
    \label{fig:drag_enhancement}
\end{figure}

\begin{align}
    \frac{dr}{dt} &= v_r, \label{eq:orbit_r}\\
    \frac{d\phi}{dt} &= \frac{v_\phi}{r}, \label{eq:orbit_phi}\\
    M_{\rm NS}\frac{dv_r}{dt} &= \frac{M_{\rm NS}v_\phi^2}{r} -\frac{Gm(r)M_{\rm NS}}{r^2} + F_{\rm net}\frac{v_r}{v}, \label{eq:orbit_vr}\\
    M_{\rm NS}\frac{dv_\phi}{dt} &= -\frac{M_{\rm NS}v_rv_\phi}{r} + F_{\rm net}\frac{v_\phi}{v}, \label{eq:orbit_vphi}
\end{align}
where $m(r)$ is the mass included within radius $r$ and $v=(v_r^2+v_\phi^2)^{1/2}$. We interpolate/extrapolate the simulation-calibrated, time-averaged force correction factor $F_{\rm net}/F_{\rm BHL}$ in $\log r$ and obtain the $F_{\rm net}$ by multiplying the factor with the BHL momentum-accretion estimate evaluated at each point. For comparison, we also calculate the orbital evolution with the canonical BHL momentum-accretion prescription for the net projected force.

The results below should be taken with caution. There are many limitations in our simulations: they are local wind-tunnel calculations; we assumed that the flow is axisymmetric and that the incoming flow is uniform, ignoring the density gradient in the stellar envelope; we neglected also the Coriolis force and the residual force after cancellation between the centrifugal and gravitational forces; most importantly the cumulative change of the envelope structure by the spiral-in of NS, and its gravitational backreaction to NS are not taken into account. It should be also mentioned that previous wind-tunnel studies found that the density gradient could reduce the accretion rate by 1--2 orders of magnitude and modify the force exerted on NS by a factor of a few \citep{2015ApJ...798L..19M,2015ApJ...803...41M,2017ApJ...838...56M}. Hence the orbital calculations presented in this section should not be regarded as quantitative predictions of CE evolution. They are intended only as illustrative estimates of the possible impact of the local force enhancement.

Figure~\ref{fig:drag_enhancement} presents the results. For the BHL momentum-accretion prescription (black curve), the orbital radius decreases only moderately over $40$\,yr. In the illustrative integration with the time-averaged simulation-calibrated force (red curve), on the other hand, the orbital evolution is much faster, and the trajectory reaches substantially smaller radii within $4$\,yr. The difference between the two curves demonstrates the possible dynamical importance of the net-force enhancement. It is mentioned again, however, that these calculations are meant for illustrative estimates rather than quantitative predictions.

We also find that the time-averaged net projected force $F_{\rm net}$ is not necessarily decelerating in our model survey, depending on the physical conditions surrounding the NS in the RSG envelope (Section~\ref{sec:4.3}). The direction of $F_{\rm net}$ is determined by the cancellation between $F_{\mathrm{pre},z}$ and $F_{\mathrm{dyn},z}$, both of which are sensitive to the inner shock structure. The nested-shock morphology varies with the envelope conditions in a non-trivial way, and therefore the resulting time-averaged sign does not follow a monotonic trend with $\rho_\infty$ or $R_a$: three models---\texttt{$\rho$1e-2}, \texttt{$\rho$3e-8}, and \texttt{$\rho$1e-8}---yield drag ($F_{\rm net}<0$) as expected in the standard BHL momentum-accretion picture, while the remaining models yield thrust ($F_{\rm net}>0$). Whether these positive values persist at higher numerical resolutions remains an open problem.

The positive net projected force, if it occurs, will accelerate rather than decelerate the NS. Then the NS will spiral out in the envelope. It should be stressed, however, that it is one thing that there is such a possibility but it is quite another whether such a situation is really encountered by the NS in the CE phase. As mentioned above, the two lowest-density models (\texttt{$\rho$3e-8} and \texttt{$\rho$1e-8}), which correspond to the outermost part of the RSG envelope, both yield negative time-averaged  net projected force, $F_{\rm net}<0$.  The NS hence spirals in at the beginning as it should, since CE could not occur otherwise. Our discrete model sequence suggests that $F_{\rm net}$ may change the sign at smaller radii. Although it is tentative and the existence and location of a radius, at which $F_{\rm net}$ vanishes, remain uncertain, if such a radius does exist, a naive expectation is that the NS does not spiral in further beyond this radius, meaning that the acceleration of NS will not be realized. This is admittedly a speculation for the moment, though, and further investigations are certainly needed.

If we nevertheless infer the radius from our model sequence, it is in the range of $a\sim40$--$70\,R_\odot$ or $\log(a/R_\odot)\sim 1.6$--1.8. If one tentatively identifies this radius with the final separation in the $\alpha$-formalism, it corresponds to $\alpha_{\rm CE}\sim0.2$--$0.4$ for $\alpha_{\rm th}=0.5$. Note again that these values should not be taken seriously, since they are obtained under the highly speculative assumption.

How and to what extent the missing physics listed above change the net projected force are unknown. In addition, magnetic fields \citep{2023ApJ...950...31K} and accretor-launched jets/outflows, if present, will further affect accretion and feed back on the envelope \citep{2017MNRAS.470.2929M,2019MNRAS.482.3646L,2019MNRAS.488.5615S,2020MNRAS.497.2057L,2022MNRAS.514.3212H}. These effects should be clarified in future works.

\subsection{Prospects for neutrino detection} \label{sec:5.2}

Another interesting question is whether the NS produces a detectable level of neutrinos. We are also concerned with their importance for cooling. The physical situation is similar to those considered for hypercritical accretion flows in TZO-like systems: the post-shock gas near the NS becomes hot and dense enough to release a non-negligible fraction of the accretion energy in neutrinos \citep{1991ApJ...376..234H,1993ApJ...411L..33C,2024ApJ...960..116A,2025ApJ...984L...2M,2026arXiv260423503C}.

To assess this possibility, we evaluate the neutrino emissivity $q_\nu(\rho,T)$ for three realistic models, \texttt{$\rho$1e-2}, \texttt{$\rho$1e-1}, and \texttt{$\rho$1e0}, which correspond to the inner region of the RSG envelope. We employ the fitting formulae of \citet{1996ApJS..102..411I} for thermal neutrino production processes. In the $(\rho,T)$ range covered by these models, the total emissivity is dominated by electron--positron pair annihilation. The event rates for representative detectors are then calculated to obtain a steady-search estimate as follows. The total neutrino luminosity and characteristic neutrino energy are estimated as
\begin{align}
    L_\nu &= \int q_\nu\, dV, \\
    \bar{E}_\nu &= \frac{\int \bar{\varepsilon}_{\nu}\, q_\nu\, dV}{\int q_\nu\, dV},
\end{align}
where $\bar{\varepsilon}_{\nu} = 3.15 k_B T$ is the average energy of neutrinos produced at a place with temperature $T$. The neutrino number flux at 10 kpc from the source is approximated as a function of energy by the following exponential form:
\begin{equation}
    \Phi_{10}(E_\nu) = \frac{L_\nu}{4\pi (10\,\mathrm{kpc})^2 \bar{E}_\nu^2}\exp\left(-\frac{E_\nu}{\bar{E}_\nu}\right),
\end{equation}
which satisfies $\int E_\nu \Phi_{10}(E_\nu)\,dE_\nu = L_\nu/[4\pi(10\,\mathrm{kpc})^2]$. The event rate is then obtained as
\begin{align}
    R_{10\,\mathrm{kpc}} &= \epsilon N_p \int_{E_{\min}}^{E_{\max}} \Phi_{10}(E_\nu)\,\sigma(E_\nu)\, dE_\nu,
\end{align}
where $\epsilon$, $N_p$, and $[E_{\min},E_{\max}]$ are the detection efficiency, number of target protons, and energy window, respectively; $\sigma$ is the cross section of the reaction employed for detection. We evaluate $\sigma$ with the following approximate form for inverse beta decay, $\sigma(E_\nu)=9.52\times10^{-44}(E_\nu-E_{\rm th})^2\,\mathrm{cm^2}$ above threshold $E_{\rm th}=1.806\,\mathrm{MeV}$, and set it to zero below threshold.

Following \citet{2025ApJ...984L...2M}, we consider steady searches with the water Cherenkov detectors Super-Kamiokande (SK) and Hyper-Kamiokande (HK), as well as with the liquid-scintillator detector JUNO. We compute $N_p$ from the fiducial detector mass and composition: $N_p=(2/18)M_{\rm fid}N_{\rm A}$ for SK and HK, and $N_p=0.12\,M_{\rm fid}N_{\rm A}$ for JUNO, where $N_{\rm A}$ is Avogadro's number. The detector parameters adopted in this analysis are summarized in Table~\ref{tab:nu_detectors}. For the background normalisation, we follow \citet{2025ApJ...984L...2M}: the HK and SK values are taken from \citet{2021PhRvD.104l2002A}, and the JUNO value from \citet{2022JCAP...10..033A}.

\begin{table}
    \centering
    \caption{Detector parameters adopted for the estimate in Section~\ref{sec:5.2}. See the text for the definitions of the symbols.}
    \label{tab:nu_detectors}
    \resizebox{\columnwidth}{!}{
    \begin{tabular}{lccccc}
        \hline
        Detector & $M_{\rm fid}$ ($\mathrm{kt}$) & $\epsilon$ & $[E_{\min},E_{\max}]$ (MeV) & $N_p$ & $r_{\rm bg}$ (Hz) \\
        \hline
        HK   & 187.0 & 0.55 & 9.3--31.3  & $1.25\times10^{34}$ & $1.87\times10^{-6}$ \\
        SK   & 22.5  & 0.55 & 9.3--31.3  & $1.50\times10^{33}$ & $2.25\times10^{-7}$ \\
        JUNO & 17.0  & 0.80 & 12.0--30.0 & $1.44\times10^{33}$ & $1.40\times10^{-7}$ \\
        \hline
    \end{tabular}}
\end{table}

For a source at distance $D$, the expected signal $(S)$ and background counts $(B)$ over an integration time $\tau$ are given, respectively, as
\begin{align}
    S(D) &= R_{10\,\mathrm{kpc}}\left(\frac{10\,\mathrm{kpc}}{D}\right)^2 \tau, \\
    B &= r_{\rm bg}\tau,
\end{align}
where $r_{\rm bg}$ is the total background rate for each detector in the same energy window (Table~\ref{tab:nu_detectors}). Detectability is then evaluated with the counting significance $S/\sqrt{S+B}$. The detection reach is the largest $D$ that satisfies $S/\sqrt{S+B}>5$ for a given $\tau$.

\begin{figure}
    \centering
    \includegraphics[width=\linewidth]{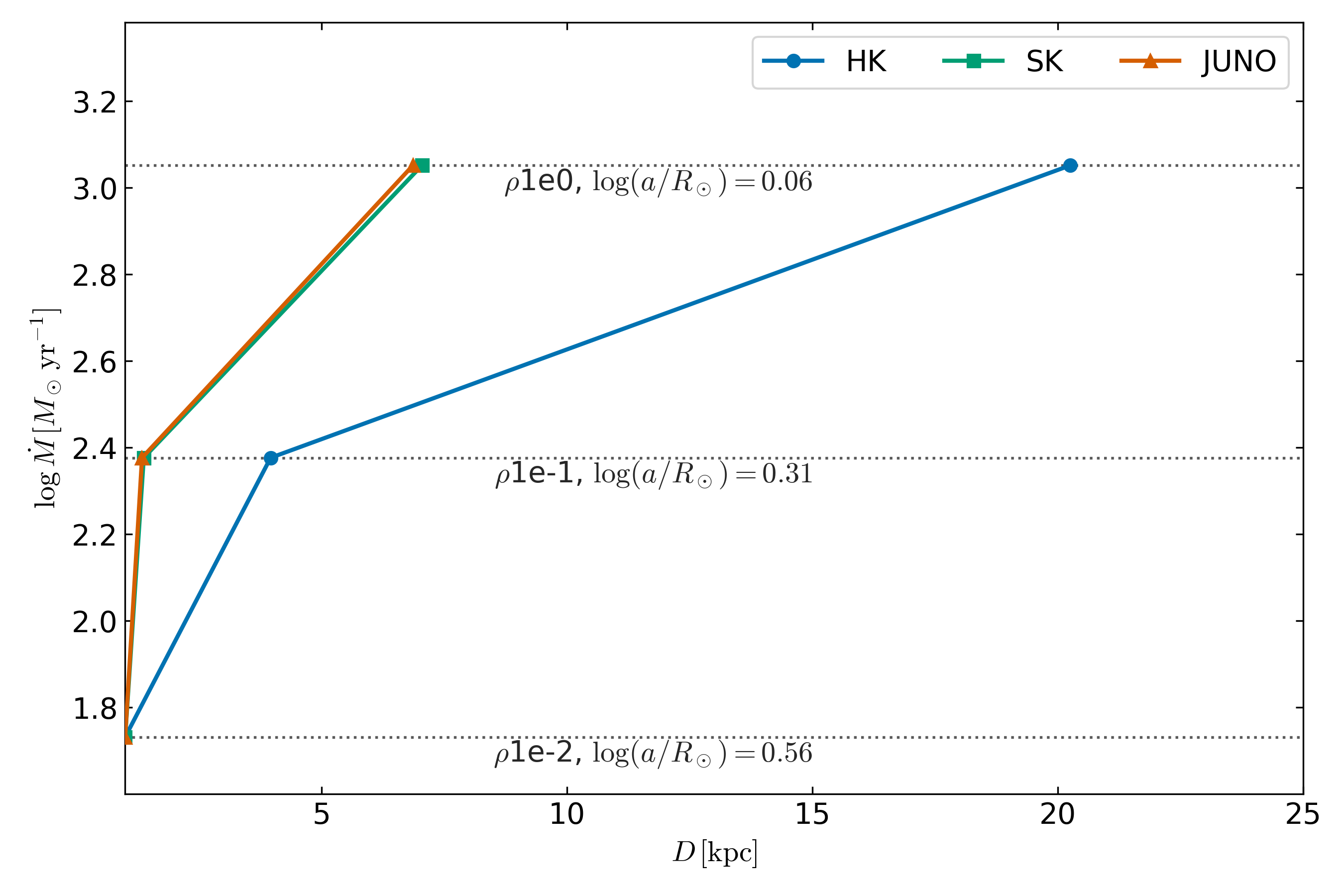}
    \caption{Detection reach inferred from the estimate in Section~\ref{sec:5.2}. The three models, \texttt{$\rho$1e0}, \texttt{$\rho$1e-1}, and \texttt{$\rho$1e-2} considered here correspond to the points in the innermost region in the RSG envelope. The horizontal dotted lines mark the accretion rates for these models with their orbital separations also attached. Different colours denote the detectors as given in the legend. For each detector, the symbol gives the maximum distance $D$ that satisfies $S/\sqrt{S+B}>5$ for each model. We assume a 1-day integration and $S$ and $B$ are the expected signal and background counts.}
    \label{fig:nu_detectability}
\end{figure}

For these three models, we estimate the gravitational accretion power as
\begin{equation}
    \dot{E}_{\rm grav}\equiv \frac{G M_{\rm NS}\dot{M}}{R_{\rm NS}}.
\end{equation}

\begin{table}
    \centering
    \caption{Gravitational accretion-power estimates $\dot{E}_{\rm grav}$ and neutrino luminosities $L_\nu$ for the three models, \texttt{$\rho$1e0}, \texttt{$\rho$1e-1}, and \texttt{$\rho$1e-2}.}
    \label{tab:nu_cooling}
    \resizebox{\columnwidth}{!}{
    \begin{tabular}{lccc}
        \hline
        Model & $\dot{E}_{\rm grav}$ [erg/s] & $L_\nu$ [erg/s] & $L_\nu / \dot{E}_{\rm grav}$ \\
        \hline
        \texttt{$\rho$1e0}  & $1.47\times10^{49}$ & $5.01\times10^{45}$ & $3.41\times10^{-4}$ \\
        \texttt{$\rho$1e-1} & $3.15\times10^{48}$ & $1.53\times10^{44}$ & $4.86\times10^{-5}$ \\
        \texttt{$\rho$1e-2} & $7.14\times10^{47}$ & $5.40\times10^{42}$ & $7.56\times10^{-6}$ \\
        \hline
    \end{tabular}}
\end{table}

Figure~\ref{fig:nu_detectability} shows the resulting detection reach for $\tau = 1$ day. Among the three models considered here, only the innermost case, \texttt{$\rho$1e0}, corresponding to $\log(a/R_\odot)=0.06$, reaches 10 kpc for HK, with a maximum distance of $20.3$ kpc, whereas SK and JUNO reach only $7.0$ and $6.9$ kpc, respectively. The two outer models, \texttt{$\rho$1e-1} and \texttt{$\rho$1e-2}, give detection reach smaller than a few kpc for all three detectors. Table~\ref{tab:nu_cooling} shows the corresponding values of $\dot{E}_{\rm grav}$ and the time-averaged values of $L_\nu$. In all three models, the ratio $L_\nu / \dot{E}_{\rm grav} \ll 1$, indicating that neutrino cooling removes only a tiny fraction of the gravitational accretion power and is unlikely to modify the flow dynamics appreciably. Note that, under the present force prescription, the illustrative trajectory in Section~\ref{sec:5.1} does not reach such deep regions of the RSG envelope. Therefore, under this prescription, the neutrino emission appears unlikely to be important observationally. Since the spiral-in depends sensitively on the gas force, however, this conclusion should be revisited in future global calculations.

%-----------------------------------------------------------------
\section{Conclusions} \label{sec:6}

We performed axisymmetric general-relativistic hydrodynamical simulations of accretion flow onto an NS engulfed in a red-supergiant envelope, resolving the accretion-flow structure self-consistently from the NS surface to the accretion radius. To cope with the extreme dynamic range ($R_a/R_{\rm NS}\sim 10^{4}$--$10^{7}$), we employed a multi-layer domain-decomposition strategy, enforcing consistency across overlapping regions between adjacent layers. The outer boundary conditions were anchored to the MESA-produced realistic models at the onset of helium-core burning, and the gas thermodynamics were calculated with the Helmholtz EOS. A survey of 10 models, spanning $\rho_\infty = 1$--$10^{-8}\,\mathrm{g\,cm^{-3}}$, was conducted to cover the range of physical conditions encountered during the spiral-in phase.

Our principal findings are as follows.

\begin{enumerate}
\item \textit{Nested-shock morphology}:
The accretion flow generically developed a nested-shock structure, in which an outer bow shock and one or more inner shocks that changed directions alternately coexisted. This morphology arose because the post-shock gas, after being decelerated at the outer bow shock, was gravitationally re-accelerated towards the NS and had to be decelerated again before reaching the surface. The formation of the inner shocks was facilitated by the high compressibility of the Helmholtz EOS ($\gamma \lesssim 4/3$) in the vicinity of the NS (Sections~\ref{sec:4.1} and \ref{sec:4.2}).

\item \textit{Accretion rate and net projected force determined differently}:
The time-averaged mass accretion rate remained within a factor of about 2 of the BHL prediction across the entire model set surveyed, with $\dot{M}/\dot{M}_{\rm BHL}\simeq 1.3$--$2.1$. In contrast, the magnitude of the time-averaged net projected force exceeded the BHL momentum-accretion estimate by one to two orders of magnitude. This finding indicates that they were determined by different physics; in particular, the net projected force was not governed by momentum accretion as envisaged in the BHL theory, but was instead controlled by the non-spherical pressure and dynamical gravity from the wake in the inner region, both of which are absent in the BHL momentum-accretion estimate (Section~\ref{sec:4.3}).

\item \textit{Thrust in some models}:
In the standard BHL momentum-accretion picture, the drag force always decelerates the NS. Some of our models produced positive net projected forces, $F_{\rm net}>0$ instead, which we referred to as thrust. The direction of the net projected force was determined by a near-cancellation between the pressure and dynamical-gravity contributions, both of which were individually large but had opposite signs and depended rather sensitively on the inner shock structure. The nested-shock morphology varied with the sampled physical conditions in a non-trivial way, and therefore the time-averaged net projected force was not necessarily negative. With our sign convention, $F_{\rm net}<0$ corresponds to drag, whereas $F_{\rm net}>0$ corresponds to thrust. The resolution and inner-boundary tests showed that the individual components of the net projected force are not fully converged. These tests were conducted only for \texttt{test-Heos}, which had $F_{\rm net}<0$, and hence did not establish the numerical robustness of the results for the realistic models, particularly those with $F_{\rm net}>0$. We therefore regarded the occurrence of thrust as tentative. The condition for thrust remained an open issue (Section~\ref{sec:4.3}).

\item \textit{Implications for orbital evolution and envelope ejection}:
The robust result is the enhancement in the magnitude of the time-averaged net projected force relative to the canonical BHL momentum-accretion estimate. In the simplified semi-analytic exercise of Section~\ref{sec:5.1}, adopting this force prescription substantially modified the orbital evolution and led to faster orbital decay than in the canonical BHL momentum-accretion case. This calculation is not a prediction of the plunge time, final separation, or CE efficiency. For the physical conditions corresponding to the outermost region in the envelope and relevant to the onset of spiral-in, the time-averaged net projected force was drag as expected. At higher densities, some models yielded thrust. From our discrete model sequence, we infer that the transition occurred in the radial range of $a\sim 40$--$70\,R_\odot$. As already mentioned, however, neither the existence nor the location of such a transition was yet robustly established. If this radius range is tentatively identified with the final separation in the $\alpha$-formalism, it corresponds to $\alpha_{\rm CE}\sim0.2$--$0.4$ for $\alpha_{\rm th}=0.5$, although this identification should not be regarded as a quantitative inference from the present local simulations (Section~\ref{sec:5.1}).

\item \textit{Neutrino emissions}:
Under the present force prescription, a post-processing estimate analogous to recent studies of TZO-like systems \citep{2025ApJ...984L...2M} showed that only the innermost model in the RSG envelope produced neutrino emission that would be detectable beyond 10 kpc with HK for a 1-day integration. Under the same prescription, the semi-analytic spiral-in trajectory did not reach such deep regions, so observationally significant neutrino emission was not expected in the current scenario. Neutrino cooling was also unlikely to affect the flow dynamics. This conclusion, however, remains sensitive to the force prescription (Section~\ref{sec:5.2}).
\end{enumerate}

These results are also important for building a subgrid model of the gas force used in CEE simulations. Since the net projected force is dominated by the pressure and dynamical gravity rather than the momentum transfer, prescriptions that tie the force to the accretion rate---as is common in the BHL momentum-accretion approach---may underestimate the magnitude of the gas force. Future prescriptions should therefore be calibrated with simulations that resolve the inner-flow structure and should track the pressure and dynamical-gravity contributions with care. The auxiliary tests conducted in the appendices seem to support this interpretation: Appendix~\ref{sec:a} validates the multi-layer domain-decomposition strategy, Appendix~\ref{sec:b} quantifies a modest but non-negligible relativistic correction to the force, Appendix~\ref{sec:c} shows that the nested-shock morphology is reproduced with Athena++, and Appendix~\ref{sec:e} shows that the enhancement of $|F_{\rm net}|$ over the BHL momentum-accretion estimate persists with increasing resolution at least for model \texttt{test-Heos}. As mentioned repeatedly above, the numerical robustness of some results for the realistic models, particularly those with $F_{\rm net}>0$, remains to be established.

The current study has some limitations that warrant future work. Firstly, we treated the RSG envelope as a fixed, unperturbed background and ignored the cumulative modification of the envelope by the spiral-in of the NS and its gravitational backreaction. Incorporating these effects may even change the estimated net projected force qualitatively. Secondly, our simulations are axisymmetric. This assumption is not fully justified, since the accretion radius is only moderately smaller than the local density scaleheight. Three-dimensional (3D) computations are therefore needed. In addition, we neglected the density gradient in the incoming wind \citep{2026arXiv260104188G}, the Coriolis force and the residual of the cancellation between the centrifugal and gravitational forces, and possible nuclear burning and associated nucleosynthesis \citep{2026ApJ...997...88A,2026arXiv260306464H}. We also neglected magnetic fields, which may affect the near-surface accretion flow, as well as accretor-launched jets/outflows, which could regulate accretion and feed back on the common-envelope evolution \citep{2017MNRAS.470.2929M,2019MNRAS.482.3646L,2019MNRAS.488.5615S,2020MNRAS.497.2057L,2022MNRAS.514.3212H,2023ApJ...950...31K}. Gravitational-wave emission should also be evaluated quantitatively \citep{2020MNRAS.493.4861G,2021ApJ...919..128R}.

%-----------------------------------------------------------------
\section*{Acknowledgements}

D.S. thanks Yudai Suwa, Akihiro Inoue, and Kosuke Mizutani for valuable discussions during his visit to Suwa's group. He also thanks Ryosuke Hirai, Damien Gagnier, Marco Vetter, and Mike Lau in "Binary Stars in a New Era (2025)" for helpful comments. He is also grateful to Tomoya Takiwaki for support at the Center for Computational Astrophysics (CfCA). Numerical computations for this work were carried out on the Cray XC50 and XD2000 systems at CfCA, National Astronomical Observatory of Japan, and on Yukawa-21 at YITP in Kyoto University. This work also used high-performance computing resources on Wisteria/BDEC-01 provided by JCAHPC through the HPCI System Research Project No. hp260198.
R.A. is supported by JSPS KAKENHI Grant 24K00632, 26K17158, and the Waseda University Grant for Special Research Projects (Project number: 2026C-742).
S.Y. is supported by JSPS KAKENHI Grant 25K01006 and the Waseda University Grant for Special Research Projects (Project number: 2025C-136).

%We are grateful to the members of our laboratory for their support, especially Wakana Iwakami, Misa Ogata, Akira Ito, and Narihiro Yamamoto.

\section*{Data availability}

The data underlying this article will be shared on reasonable request to the corresponding author.

\bibliographystyle{mnras}
\bibliography{complete_manuscript_file}

\appendix
%-----------------------------------------------------------------
\section{Necessity and validation of the multi-layer domain-decomposition} \label{sec:a}

In Section~\ref{sec:3.2}, we introduced the multi-layer domain-decomposition strategy used to connect the NS surface to the accretion radius. Here we explain why a single-domain calculation is impractical for the present problem and verify, with a controlled test, that the multi-layer treatment does not introduce a measurable bias in the diagnostic quantities relevant for this work.

\subsection{Why the multi-layer domain-decomposition is necessary} \label{sec:a1}

The main obstacle to a single-domain calculation is not the number of grid cells alone, but the extreme disparity of time-scales between the NS surface and the accretion radius. In an explicit Godunov-type scheme, the global Courant--Friedrichs--Lewy (CFL) time step is set by the smallest cells near the NS surface. Consequently, even the outer flow at $r\sim R_a$, whose characteristic advection time is much longer, must be evolved with the tiny time step imposed by the innermost cells.

To see this, consider a logarithmic grid with constant fractional spacing $\Delta r / r \simeq \epsilon$. The global CFL time step is then approximately
\begin{equation}
\Delta t_{\rm single} \sim \frac{\epsilon R_{\rm NS}}{c_s(R_{\rm NS})},
\end{equation}
where $c_s(R_{\rm NS})$ is the characteristic signal speed near the NS surface. To reach a quasi-steady state over the full radial extent, the outer flow must be evolved for a physical time comparable to the flow-crossing time,
\begin{equation}
t_{\rm cross} \sim \frac{R_a}{v_\infty}.
\end{equation}
The required number of time steps is therefore
\begin{equation}
N_{\rm step} \sim \frac{t_{\rm cross}}{\Delta t_{\rm single}}
\sim \frac{R_a}{\epsilon R_{\rm NS}}
\frac{c_s(R_{\rm NS})}{v_\infty}.
\end{equation}
For representative values $R_a/R_{\rm NS}\sim 10^7$, $\epsilon=0.05$, $c_s(R_{\rm NS})\sim 0.1c$, and $v_\infty\sim 10^{-3}c$, we obtain
\begin{equation}
N_{\rm step}\sim 2\times 10^{10}.
\end{equation}
Thus, the prohibitive cost of a single-domain calculation arises primarily from the global time-stepping constraint, rather than from the number of grid cells itself.

The multi-layer domain-decomposition alleviates this bottleneck because each layer is evolved with its own local time step and only over its own radial extent, while the global solution is recovered by matching adjacent layers in the overlapping regions as described in Section~\ref{sec:3.2}. Each layer $\ell$ has its own inner radius $R_{\rm in}^{(\ell)}$ and therefore its own minimum cell size, so that the local CFL time step scales as
\begin{equation}
\Delta t_{(\ell)} \sim \frac{\epsilon R_{\rm in}^{(\ell)}}{c_{\rm char}^{(\ell)}},
\end{equation}
where $c_{\rm char}^{(\ell)}$ denotes the characteristic signal speed in that layer. Outer layers therefore permit much larger time steps than the innermost one, because both $R_{\rm in}^{(\ell)}$ is larger and the local characteristic speed is smaller. The total computational cost is thereby reduced by many orders of magnitude relative to a single-domain run while maintaining consistency across the entire radial range, because no outer layer is forced to evolve with the innermost CFL time step.

\subsection{Validation: one-layer vs.\ multi-layer for the test model} \label{sec:a2}

We now verify that the multi-layer domain-decomposition does not introduce measurable bias by applying an artificial two-layer split to a configuration that is otherwise tractable in a single domain, namely the \texttt{test-Heos} model. Table~\ref{tab:convergence_layers} summarizes the adopted two-layer configuration. We use the same convergence criterion as described in Section~\ref{sec:3.2} and adopt the same sign convention as in Section~\ref{sec:3.3}: $F_{\rm net}<0$ corresponds to a force that decelerates the NS. Table~\ref{tab:convergence_check} then compares the converged diagnostic values obtained with the one- and two-layer setups, in the same format as Table~\ref{tab:results_summary}.

\begin{table*}
\centering
\caption{Artificial two-layer split adopted for the Appendix~\ref{sec:a2} validation test with \texttt{test-Heos}. Here $N_r^{\mathrm{all}}$ denotes the number of radial cells across the full computational domain.}
\label{tab:convergence_layers}
\begin{tabular}{cccccccc}
\hline
model & $n_\mathrm{layer}$ &
$R_\mathrm{in}^{2^\mathrm{nd}} \text{--} R_\mathrm{out}^{2^\mathrm{nd}}\,(\mathrm{cm})$ & $N_r^{2^\mathrm{nd}}$ &
$R_\mathrm{in}^{1^\mathrm{st}} \text{--} R_\mathrm{out}^{1^\mathrm{st}}\,(\mathrm{cm})$ & $N_r^{1^\mathrm{st}}$ &
$N_r^{\mathrm{all}}$ & $N_\theta$ \\
\hline
\texttt{test-Heos} (two-layer) & 2 &
$1.15 \times 10^6 \text{--} 7 \times 10^8$ & $140$ &
$3\times 10^8 \text{--} 3\times 10^9$ & $42$ &
$168$ & $48$ \\
\hline
\end{tabular}
\end{table*}

\begin{table*}
\centering
\caption{Validation of the multi-layer domain-decomposition strategy using the \texttt{test-Heos} model. The listed diagnostics follow the same definitions as in Table~\ref{tab:results_summary}.}
\label{tab:convergence_check}
\begin{tabular}{lcccccc}
\hline
Configuration &
$\dot{M}$ (g/s) &
$F_{\mathrm{mom},z}$ (dyn) &
$F_{\mathrm{pre},z}$ (dyn) &
$F_{\mathrm{dyn},z}$ (dyn) &
$F_{\mathrm{net}}$ (dyn) &
$\dot{E}_{\rm out}$ (erg/s) \\
\hline
Single layer & $5.0 \times 10^{29} (1.6)$ & $-5.4 \times 10^{37}$ & $-2.3 \times 10^{39}$ & $3.0 \times 10^{38}$ & $-2.1 \times 10^{39} (-6.5)$ & $5.1 \times 10^{47} (1.6)$ \\
Two layers   & $4.6 \times 10^{29} (1.5)$ & $-2.2 \times 10^{37}$ & $-1.1 \times 10^{39}$ & $-8.0 \times 10^{38}$ & $-2.0 \times 10^{39} (-6.3)$ & $5.8 \times 10^{47} (1.8)$ \\
\hline
\end{tabular}
\end{table*}

The accretion rate, net projected force, and energy deposition rate at the outer boundary agree closely between the two setups. The individual force components differ more noticeably, but their sums remain similar. This confirms that the multi-layer domain-decomposition does not introduce a systematic bias in the total diagnostics relevant for the conclusions of this work.

%-----------------------------------------------------------------
\section{Relativistic effects} \label{sec:b}

We next assess how strongly relativistic gravity affects the inferred net projected force. Although the outer flow is approximately Newtonian, the gas near the NS surface samples a much deeper potential, and relativistic corrections may therefore be non-negligible in the region that is most relevant to the force. To quantify this effect, we compare our default Schwarzschild calculation with an otherwise identical Newtonian-limit run.

In ordinary spherical coordinates under axisymmetry, the Newtonian conservation laws are
\begin{align}
&\partial_t \rho
+ \frac{1}{r^2}\partial_r \left(r^2 \rho v_r\right)
+ \frac{1}{r\sin\theta}\partial_\theta \left(\sin\theta\, \rho v_\theta\right) = 0,
\label{eq_massconserv_newt}\\[4pt]
&\partial_t (\rho v_r)
+ \frac{1}{r^2}\partial_r \left[r^2 \left(\rho v_r^2 + P\right)\right]
+ \frac{1}{r\sin\theta}\partial_\theta \left(\sin\theta\, \rho v_r v_\theta\right)
\nonumber\\
&\qquad = -\rho \frac{GM_{\rm NS}}{r^2}
+ \frac{2P}{r}
+ \rho \frac{v_\theta^2}{r},
\label{eq_Euler_newt1}\\[4pt]
&\partial_t (\rho v_\theta)
+ \frac{1}{r^2}\partial_r \left(r^2 \rho v_r v_\theta\right)
+ \frac{1}{r\sin\theta}\partial_\theta \left[\sin\theta \left(\rho v_\theta^2 + P\right)\right]
\nonumber\\
&\qquad = \frac{P\cot\theta}{r}
- \rho \frac{v_r v_\theta}{r},
\label{eq_Euler_newt2}\\[4pt]
&\partial_t e
+ \frac{1}{r^2}\partial_r \left[r^2 \left(e + P\right) v_r \right]
\nonumber\\
&\qquad + \frac{1}{r\sin\theta}\partial_\theta \left[\sin\theta \left(e + P\right) v_\theta \right]
= -\rho v_r \frac{GM_{\rm NS}}{r^2},
\label{eq_energyconserv_newt}
\end{align}
Here $v_r$ and $v_\theta$ are the radial and polar velocity components, and the total energy density is
\begin{equation}
e = \frac{1}{2}\rho \left(v_r^2 + v_\theta^2\right) + \rho \varepsilon.
\end{equation}
where $\varepsilon$ is the specific internal energy.
We use this Newtonian formulation to construct a counterpart of the \texttt{test-Heos} model and compare the resulting force with the GR result. In both cases, $F_{\rm net}$ is computed using the diagnostic definitions in Section~\ref{sec:3.3} with the same sign convention ($F_{\rm net}<0$: decelerating).

Table~\ref{tab:relativistic_comparison} summarizes the comparison for \texttt{test-Heos}. In addition to the accretion rate and net projected force, we list its individual components and the energy deposition rate. This allows us to identify which element in the force balance is affected most by the relativistic treatment.

\begin{table*}
\centering
\caption{Mass accretion rates, force components, and energy deposition rates at the outer boundary for the relativistic and Newtonian calculations of \texttt{test-Heos}.}
\label{tab:relativistic_comparison}
\begin{tabular}{lcccccc}
\hline
Method &
$\dot{M}$ (g/s) &
$F_{\mathrm{mom},z}$ (dyn) &
$F_{\mathrm{pre},z}$ (dyn) &
$F_{\mathrm{dyn},z}$ (dyn) &
$F_{\mathrm{net}}$ (dyn) &
$\dot{E}_{\rm out}$ (erg/s) \\
\hline
Newtonian & $3.5 \times 10^{29}\,(1.1)$ & $4.0 \times 10^{37}$ & $5.9 \times 10^{38}$ & $-2.1 \times 10^{39}$ & $-1.5 \times 10^{39}\,(-4.7)$ & $5.1 \times 10^{47}\,(1.6)$ \\
GR        & $5.0 \times 10^{29}\,(1.6)$ & $-5.4 \times 10^{37}$ & $-2.3 \times 10^{39}$ & $3.0 \times 10^{38}$ & $-2.1 \times 10^{39}\,(-6.5)$ & $5.1 \times 10^{47}\,(1.6)$ \\
\hline
\end{tabular}
\\
\raggedright
\textit{Note.} The numbers in parentheses show the values scaled by the corresponding BHL momentum-accretion estimates, as in Table~\ref{tab:results_summary}.
\end{table*}

The comparison shows that relativistic corrections are modest but not negligible, and they arise primarily from the near-surface region where the lapse $\alpha(r)=\sqrt{1-2GM_{\rm NS}/(rc^2)}$ and its gradients directly affect the momentum and energy balance (Section~\ref{sec:3.1}). We therefore retain the GR treatment throughout this work to ensure consistent near-surface dynamics even when the outer flow is approximately Newtonian.

%-----------------------------------------------------------------
\section{Code-independence of the nested-shock structure} \label{sec:c}

We next verify that the nested-shock morphology found in Section~\ref{sec:4.1} is not an artefact of our code implementation. For this purpose, we performed an independent Newtonian calculation with Athena++ \citep{2020ApJS..249....4S} for the \texttt{test-$\gamma$4/3} model, which is the simplest test model that exhibits a clear nested-shock structure.

\begin{figure*}
    \centering
    \includegraphics[width=\linewidth]{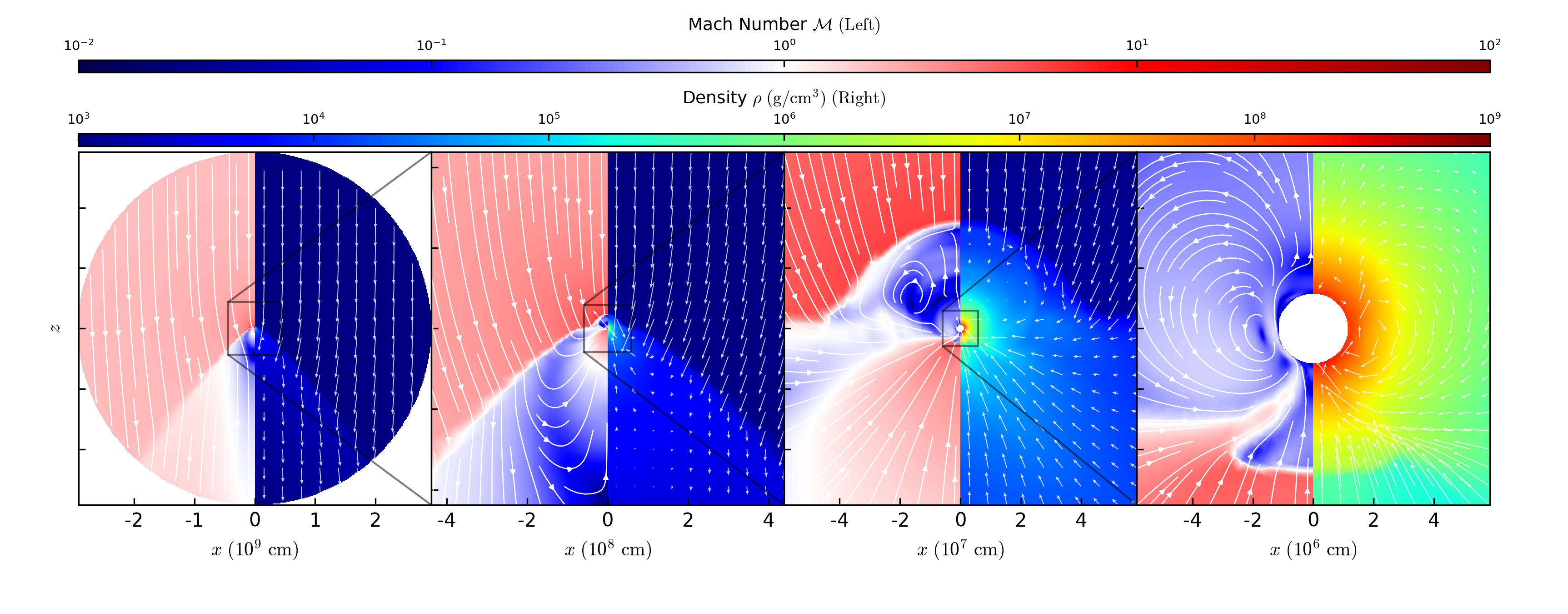}
    \caption{Athena++ result for the Newtonian \texttt{test-$\gamma$4/3} model. The panel layout follows that of the bottom row of Fig.~\ref{fig:test_gamma}: the left half of each panel shows the Mach number with streamlines, while the right half shows the density with velocity vectors. The flow develops the same qualitative nested-shock morphology found with our code, namely an outer bow shock and an inner, oppositely oriented shock.}
    \label{fig:athena_gamma43}
\end{figure*}

The purpose of this comparison is not a detailed code-to-code calibration, but a qualitative check that an independent hydrodynamical solver recovers the same shock morphology. Figure~\ref{fig:athena_gamma43} shows the corresponding Athena++ result for the Newtonian \texttt{test-$\gamma$4/3} model: an outer bow shock forms first, and the post-shock gas is then gravitationally re-accelerated and decelerated again by an inner, oppositely oriented shock, as in our calculation. This agreement indicates that the nested-shock structure is not specific to our numerical implementation and is robust at the qualitative level relevant for the main argument of this paper.

%-----------------------------------------------------------------
\section{Sensitivity to inner boundary conditions} \label{sec:d}

The NS surface is represented by an inner boundary at $r=R_{\rm NS}$, as described in Section~\ref{sec:3.2}. Since the innermost shocked region remains oscillatory, no unique steady boundary state exists there. We therefore test how sensitive the inferred quasi-steady drag is to several reasonable inner-boundary closures.

We consider five boundary treatments:

\begin{figure*}
    \centering
    \includegraphics[width=\linewidth]{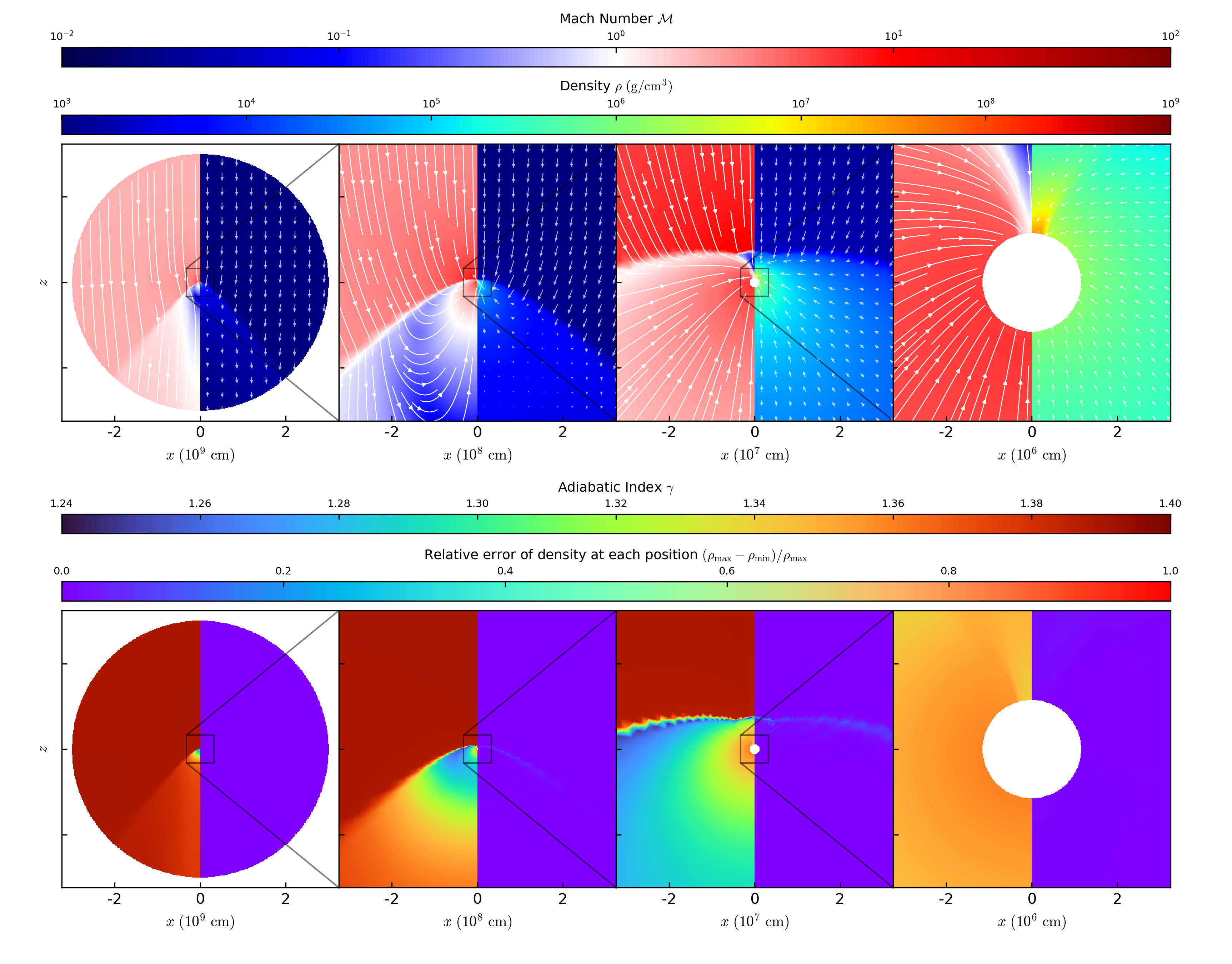}
    \caption{Same as Fig.~\ref{fig:test_heos}, but for the BH-like sink boundary. In this case, the flow can pass through the inner boundary supersonically, so the inner, oppositely oriented shock does not form. The lower panels also show that the strong time variability associated with the instability discussed in Section~\ref{sec:4.1} is almost absent.}
    \label{fig:test_inner_sink}
\end{figure*}

\begin{enumerate}
\item \textit{Baseline (main text)}: impose $v_r=0$ at $R_{\rm NS}$ and populate ghost zones by copying $\rho$ and $T$ from the innermost active cells.

\item \textit{Dynamical time averaged}: impose $v_r=0$ at $R_{\rm NS}$, but replace the instantaneous copy of $\rho$ and $T$ by time-averaged values. Specifically, we compute running averages $\langle \rho \rangle$ and $\langle T \rangle$ at the innermost active cells over a window comparable to the dynamical time to $r\simeq 10^{9}\,{\rm cm}$,
\begin{equation}
t_{\rm dyn}(10^{9}\,{\rm cm}) \equiv 10^9/c_s(R_{\rm NS}) \approx 1\,{\rm s},
\end{equation}
which is also comparable to the characteristic oscillation time-scale of the inner shock in this test. The ghost zones are then filled using $\langle \rho \rangle$ and $\langle T \rangle$ rather than the instantaneous values. This variant mimics a smoothed boundary without requiring an arbitrary constant choice of $(\rho,T)$.

\item \textit{Time-cumulative}: impose $v_r=0$ at $R_{\rm NS}$, but replace the instantaneous copy of $\rho$ and $T$ by cumulative averages measured from the beginning of the simulation.

\item \textit{Sink}: construct the ghost zones by extrapolating the active-mesh values inward and impose a diode condition at the inner boundary so that matter can flow into the boundary but not emerge from it. This is a BH-like sink prescription.

\item \textit{Reflective}: reflect the radial velocity ($v_r\rightarrow -v_r$) at $R_{\rm NS}$, while keeping the same copying for $\rho$ and $T$. This provides an upper-limit test of how strongly reflection could modify the inner shock and wake morphology.
\end{enumerate}

Among the NS-like closures, namely the baseline, time-averaged, time-cumulative, and reflective cases, the nested-shock structure is retained and the inferred drag varies only at a modest level. The sink boundary, however, is qualitatively different. In this case, the flow can pass through the inner boundary supersonically, so no inner, oppositely oriented shock is required and the nested-shock structure therefore does not form.

Figure~\ref{fig:test_inner_sink} shows the corresponding result for the BH-like sink boundary. This case serves as a contrast test rather than an alternative NS-surface boundary condition: because the compact object is effectively treated in a BH-like manner, the strong time variability associated with the entropic-acoustic instability is largely absent.

Table~\ref{tab:boundary_sensitivity} shows the resulting diagnostic values for \texttt{test-Heos}, including the accretion rate, individual force components, net projected force, and energy deposition rate. The individual components are more sensitive to the boundary condition than the total force, showing greater variabilities in the inner shocked region and different proportions of the momentum, pressure, and dynamical-gravity terms in the force balance. Nevertheless, the four NS-like closures give the same sign and comparable magnitudes to $F_{\rm net}$, indicating that the net projected force is controlled primarily by the global shocked flow rather than by a single boundary condition. The sink case is listed separately as a contrast case because, although its diagnostic values remain broadly comparable, it alters the qualitative flow morphology itself.

\begin{table*}
\centering
\caption{Sensitivity of the \texttt{test-Heos} results to the inner boundary treatment. The listed diagnostics follow the same definitions as in Table~\ref{tab:results_summary}. The BH-like sink boundary is shown separately as a contrast test.}
\label{tab:boundary_sensitivity}
\begin{tabular}{lcccccc}
\hline
Boundary type &
$\dot{M}$ (g/s) &
$F_{\mathrm{mom},z}$ (dyn) &
$F_{\mathrm{pre},z}$ (dyn) &
$F_{\mathrm{dyn},z}$ (dyn) &
$F_{\mathrm{net}}$ (dyn) &
$\dot{E}_{\rm out}$ (erg/s) \\
\hline
Baseline & $5.0 \times 10^{29}\,(1.6)$ & $-5.4 \times 10^{37}$ & $-2.3 \times 10^{39}$ & $3.0 \times 10^{38}$ & $-2.1 \times 10^{39}\,(-6.5)$ & $5.1 \times 10^{47}\,(1.6)$ \\
Dynamical time averaged & $4.6 \times 10^{29}\,(1.5)$ & $-2.3 \times 10^{37}$ & $-1.2 \times 10^{39}$ & $-8.4 \times 10^{38}$ & $-2.1 \times 10^{39}\,(-6.6)$ & $5.2 \times 10^{47}\,(1.6)$ \\
Time cumulative & $4.6 \times 10^{29}\,(1.5)$ & $7.5 \times 10^{37}$ & $-3.3 \times 10^{39}$ & $3.4 \times 10^{38}$ & $-2.9 \times 10^{39}\,(-9.1)$ & $5.1 \times 10^{47}\,(1.6)$ \\
Sink & $4.6 \times 10^{29}\,(1.4)$ & $4.2 \times 10^{38}$ & $-2.1 \times 10^{38}$ & $-4.4 \times 10^{39}$ & $-4.2 \times 10^{39}\,(-13.0)$ & $5.2 \times 10^{47}\,(1.6)$ \\
Reflective & $-1.3 \times 10^{27}\,(-0.004)$ & $-5.7 \times 10^{38}$ & $3.9 \times 10^{38}$ & $-1.2 \times 10^{39}$ & $-1.4 \times 10^{39}\,(-4.5)$ & $1.2 \times 10^{48}\,(3.7)$ \\
\hline
\end{tabular}
\end{table*}

The spread in the NS-like variants is small compared with the one-to-two-order-of-magnitude enhancement of $|F_{\rm net}|$ over the BHL momentum-accretion estimate found in the main model set (Section~\ref{sec:4.3}). The decomposed force components should therefore be interpreted as diagnostics of the force balance rather than as separately converged quantities. What is robust in this test with model \texttt{test-Heos} is the negative sign and order of magnitude of their sum, $F_{\rm net}$, for reasonable NS-surface boundary treatments. The numerical robustness of the results for the realistic models, particularly those with $F_{\rm net}>0$, is yet to be established.

%-----------------------------------------------------------------
\section{Resolution study} \label{sec:e}

Finally, we examine the numerical convergence of the key diagnostic quantities. We use the \texttt{test-Heos} model, which is handled with a single layer and can therefore be evolved over the full radial range at several resolutions. We refine both the radial and polar grids by approximately a factor of two at each step.

\begin{table*}
\centering
\caption{Resolution study for \texttt{test-Heos}. The listed diagnostics follow the same definitions as in Table~\ref{tab:results_summary}.}
\label{tab:convergence}
\begin{tabular}{lccccccc}
\hline
Resolution & $N_r\times N_\theta$ &
$\dot{M}$ (g/s) &
$F_{\mathrm{mom},z}$ (dyn) &
$F_{\mathrm{pre},z}$ (dyn) &
$F_{\mathrm{dyn},z}$ (dyn) &
$F_{\mathrm{net}}$ (dyn) &
$\dot{E}_{\rm out}$ (erg/s) \\
\hline
Low    & $168\times 48$  & $5.0 \times 10^{29}\,(1.6)$ & $-5.4 \times 10^{37}$ & $-2.3 \times 10^{39}$ & $3.0 \times 10^{38}$ & $-2.1 \times 10^{39}\,(-6.5)$ & $5.1 \times 10^{47}\,(1.6)$ \\
Medium & $336\times 96$  & $4.2 \times 10^{29}\,(1.3)$ & $2.2 \times 10^{37}$ & $2.3 \times 10^{39}$ & $-4.7 \times 10^{39}$ & $-2.4 \times 10^{39}\,(-7.6)$ & $4.6 \times 10^{47}\,(1.5)$ \\
High   & $676\times 192$ & $3.9 \times 10^{29}\,(1.2)$ & $4.3 \times 10^{36}$ & $1.2 \times 10^{39}$ & $-3.7 \times 10^{39}$ & $-2.5 \times 10^{39}\,(-7.9)$ & $3.9 \times 10^{47}\,(1.2)$ \\
\hline
\end{tabular}
\end{table*}

Table~\ref{tab:convergence} shows the results. As the grid is refined from $168\times 48$ to $676\times 192$, the mass accretion rate changes only modestly, indicating that $\dot{M}$ is captured reasonably well at the fiducial survey resolution and is well converged once the grid is refined further. The energy deposition rate at the outer boundary also changes moderately, by a factor of $\sim 1.3$ between the low- and high-resolution runs. The shock locations remain stable to within a few percent.

The individual force components are more precarious. Whereas the momentum-flux term is always subdominant, the pressure and dynamical-gravity terms are individually large in modulus, mostly cancel each other, and are more sensitive to resolution. In fact, the value $F_{\mathrm{pre},z}$ changes from $-2.3\times10^{39} \mathrm{dyn}$ at the lowest resolution to $2.3\times10^{39}$ and $1.2\times10^{39} \mathrm{dyn}$ at the medium and the highest resolutions, respectively; the value of $F_{\mathrm{dyn},z}$ varies from $3.0\times10^{38}$ to $-4.7\times10^{39}$ to $-3.7\times10^{39} \mathrm{dyn}$ for the same resolutions. It is hence apparent that they have not yet converged numerically; we use the individual components for a qualitative diagnosis of force balance alone.

The total net projected force is nevertheless more stable than its individual components. Across the same resolution range, $F_{\rm net}$ remains negative and varies from $-2.1\times10^{39}$ to $-2.5\times10^{39} \mathrm{dyn}$, corresponding to $|F_{\rm net}|/F_{\rm BHL}=6.5$--7.9. Although component-wise convergence has not been achieved, the negative sign of $F_{\rm net}$ and its substantial enhancement over the BHL momentum-accretion estimate are rather robust at least for \texttt{test-Heos}. We therefore draw the following conclusions from this resolution study: (i) the presence of the nested shocks is not a numerical artifact and their number is also robust, (ii) $\dot{M}$ is well converged, and (iii) the total net projected force is enhanced over the BHL momentum-accretion estimate by a factor of several, while its value is accurate only at the order-of-magnitude level at present. Note, however, that the numerical robustness of the results for the realistic models, particularly those with $F_{\rm net}>0$, is remaining to be demonstrated.

\bsp
\label{lastpage}

\end{document}